\documentclass[letterpaper, 12pt]{article}
\usepackage{jheppub}

\usepackage{graphicx}
\usepackage{bbm}
\usepackage{amsmath}
\usepackage{amssymb}
\usepackage{mathtools}
\usepackage{amsthm}
\usepackage{physics}
\usepackage{float}
\usepackage{upgreek}
\usepackage{afterpage}
\usepackage{multirow}
\usepackage{rotating}
\usepackage{epstopdf}
\usepackage{mathrsfs}
\usepackage[labelformat=simple]{subcaption}
\usepackage{calligra}
\usepackage[T1]{fontenc}  
  
\begin{document}

\begin{titlepage}

\setcounter{page}{1} \baselineskip=15.5pt \thispagestyle{empty}
{\flushright {ITP-CAS-25-088}\\}
		
\bigskip\
		
\vspace{0.4cm}
\begin{center}
%{%\fontsize{20}{28}
%{\LARGE \bfseries the\vspace{0.24cm}\\ string}
{\LARGE \bfseries Axion Mixing in the String Axiverse}
\end{center}
%\vspace{0.5cm}
\vspace{0.15cm}
			
\begin{center}
{\fontsize{14}{30}\selectfont Hai-Jun Li$^{a,b,c}$ and Yu-Feng Zhou$^{b,c,d,e}$}
\end{center}
%\vspace{0.1 cm}
\begin{center}
\vspace{0.25 cm}
\textsl{$^a$Department of Physics, Yunnan University, Kunming 650091, China}\\
\textsl{$^b$Institute of Theoretical Physics, Chinese Academy of Sciences, Beijing 100190, China}\\
\textsl{$^c$International Centre for Theoretical Physics Asia-Pacific, Beijing 100190, China}\\
\textsl{$^d$School of Physical Sciences, University of Chinese Academy of Sciences, Beijing 100049, China}\\
\textsl{$^e$School of Fundamental Physics and Mathematical Sciences, Hangzhou Institute for Advanced Study, UCAS, Hangzhou 310024, China}\\

\vspace{-0.1 cm}				
\begin{center}
{E-mail: \textcolor{blue}{\tt {lihaijun@ynu.edu.cn}}, \textcolor{blue}{\tt {yfzhou@itp.ac.cn}}}
\end{center}	
\end{center}
\vspace{0.6cm}
\noindent

String axiverse provides a fascinating and complex landscape for axion physics.
The requirement in type IIB string axiverse models necessitates at least two axions to ensure the presence of both a QCD axion candidate and an additional axion-like particle (ALP).
In this work, we study axion mass mixing and adopt a bottom-up perspective to investigate what conditions an axion model must satisfy in order to exhibit maximal mixing --- the scenario where the degree of effective mixing is maximized.
We find that maximal mixing occurs when the masses of all ALPs are smaller than the zero-temperature mass of the QCD axion, with no two ALP masses being equal, and when the decay constants of all ALPs are uniformly either smaller or larger than the decay constant of the QCD axion. 
Additionally, the transfer of axion energy density ultimately takes place only between the two axions with the closest masses. 
These findings provide critical insights into axion dynamics not only within type IIB string axiverse models but also in broader multi-axion mixing frameworks.
The potential cosmological implications of axion mass mixing are also addressed at the end.
			
\vspace{1.6cm}
			
\bigskip
\noindent\today
\end{titlepage}
			
\setcounter{tocdepth}{2}
			
\hrule
\tableofcontents			
\bigskip\medskip
\hrule
\bigskip\bigskip
%\pagebreak\
 
\newpage
\section{Introduction}%%%%%%%%%%%%%%%%%%%%%%%%%%Introduction

String theory phenomenology introduces the concept of the Axiverse --- a cosmological framework predicting not only the QCD axion in the low-energy effective field theory (EFT), but also a plenitude of ultra-light axion-like particles (ALPs) \cite{Arvanitaki:2009fg, Demirtas:2018akl, Reig:2021ipa, Demirtas:2021gsq}. 
The QCD axion, a pseudoscalar field, originates from the Peccei-Quinn (PQ) mechanism \cite{Peccei:1977hh, Peccei:1977ur}, which postulates the spontaneous breaking of a global $\rm U(1)_{PQ}$ symmetry. 
Under this PQ symmetry, the axion field $\phi$ transforms as $\phi \rightarrow \phi + 2\pi f_a$ (where $f_a$ represents the axion decay constant), dynamically resolving the strong CP problem in the Standard Model (SM) by stabilizing its potential at the CP-conserving minimum \cite{Weinberg:1977ma, Wilczek:1977pj, Kim:1979if, Shifman:1979if, Dine:1981rt, Zhitnitsky:1980tq}. 
This stabilization arises from QCD non-perturbative effects, which induce the axion potential via QCD instantons and generate a parametrically small axion mass \cite{tHooft:1976rip, tHooft:1976snw}.  

Moreover, the QCD axion serves as a compelling cold dark matter candidate. 
Its non-thermal production in the early Universe, driven by the misalignment mechanism, links the oscillation energy density of the axion field to the observed dark matter abundance \cite{Preskill:1982cy, Abbott:1982af, Dine:1982ah}.
Recent reviews further elaborate on these aspects \cite{Marsh:2015xka, DiLuzio:2020wdo, OHare:2024nmr, Hook:2018dlk}. 
For a natural initial misalignment angle $\theta_{\rm i} \sim \mathcal{O}(1)$, this mechanism constrains the QCD axion decay constant to the classical window
\begin{eqnarray}
10^{8}-10^{9}\, {\rm GeV} \lesssim f_a\lesssim 10^{11}-10^{12}\, {\rm GeV}\, .
\label{axion_classical_window}
\end{eqnarray}
The upper bound stems from avoiding the overproduction of axion dark matter, while the lower limit is derived from astrophysical probes such as neutrino burst durations in SN\,1987A \cite{Raffelt:1987yt, Turner:1987by, Mayle:1987as} and neutron star cooling rates \cite{Leinson:2014ioa, Hamaguchi:2018oqw, Buschmann:2021juv}. 

On the other hand, a plenitude of ultra-light ALPs with logarithmically hierarchical masses, which also encompass the QCD axion, predominantly arise from higher-dimensional gauge fields \cite{Witten:1984dg, Green:1984sg, Choi:2003wr}. 
In the framework of string theory \cite{Svrcek:2006yi, Conlon:2006tq}, these particles can be understood as lower-dimensional axions emerging through the process of integrating higher-dimensional gauge fields over specific cycles within the compactified extra dimensions. 
There exist model-independent axions, which remain unaffected by the detailed structure of the internal manifold. 
Conversely, there also exist axions whose decay constants are significantly influenced by the internal geometry of the compactification, giving rise to what is commonly termed model-dependent axions.
These model-independent and model-dependent axions highlight the rich interplay between higher-dimensional theories and low-energy phenomenology, thus providing a promising arena for exploring new physics beyond the SM.
For further insights into the role of axions in extra-dimensional scenarios, including recent advancements and discussions, see, for example, refs.~\cite{Gendler:2024gdo, Reece:2024wrn, Agrawal:2024ejr, Li:2024jko, Choi:2024ome}. 

It was shown that the potential of an axiverse can be realized within the framework of type IIB string compactifications \cite{Acharya:2010zx, Cicoli:2012sz, Cicoli:2011qg, Cicoli:2012cy, Broeckel:2021dpz}.
Below the compactification scale, the low-energy spectrum typically contains many ALPs, which are Kaluza-Klein (KK) zero modes of antisymmetric form fields originating from bosonic type IIB strings. 
Specifically, their number, which is linked to the topology of the internal manifold and particularly to the harmonic forms on a Calabi-Yau (CY) manifold, is often in the hundreds.
Therefore, the low-energy theory is rich in these closed string axions. 
While open string axions may also exist on a brane, they are more model-dependent, so the focus here is on closed string axions in the bulk.

Each axion is accompanied by a real scalar field known as the saxion, together forming a complex field that constitutes the lowest component of a chiral superfield. 
The saxion and axion represent the real and imaginary parts of this complex field, respectively.
Saxions are moduli that affect EFT features, and it is crucial for them to avoid mediating the unobserved fifth-forces and suffering from the cosmological moduli problem (CMP) \cite{Coughlan:1983ci, deCarlos:1993wie, Banks:1993en}.
Then the moduli stabilization mechanism, such as the Kachru-Kallosh-Linde-Trivedi (KKLT) scenario \cite{Kachru:2003aw}, gives the moduli a heavy mass, which may also induce a potential for axions that makes them too heavy.
In refs.~\cite{Cicoli:2012sz, Cicoli:2011qg, Cicoli:2012cy}, it was shown that a viable axiverse with multiple light axions can be realized in the context of the LARGE volume scenario (LVS) \cite{Balasubramanian:2005zx, Blumenhagen:2009gk}.

The type IIB string axiverse features two regimes based on the vacuum expectation value (VEV) of the tree-level superpotential $W_0$. 
When $W_0\ll 1$, a single non-perturbative effect creates a minimum for all Kähler moduli, lifting one axion while leaving the other axions massless \cite{Acharya:2010zx}.
However, in this regime, the challenges faced include identifying a rigid ample divisor and avoiding chiral intersections.
When $W_0$ is of order one, the LVS moduli stabilization mechanism results in a heavy axion from a del Pezzo divisor fixed by non-perturbative effects and multiple light axions stabilized through perturbative effects, acquiring a potential from higher-order instanton effects \cite{Balasubramanian:2005zx, Blumenhagen:2009gk, Cicoli:2008va}. 
Furthermore, some light axions could be eaten up by massive $\rm U(1)$ gauge bosons via the Stückelberg mechanism.

The LVS axiverse requires at least two axions to guarantee the presence of both a QCD axion candidate and at least one additional ALP.
Specifically, ref.~\cite{Cicoli:2012sz} outlines models featuring one QCD axion plus one ALP, two ALPs, and multiple ALPs.
In general, realizing the QCD axion requires an intermediate-scale decay constant and an anomaly coefficient of order unity. 
In the LVS axiverse, the QCD axion with a higher-scale decay constant can be diluted by the late out-of-equilibrium decay of some heavy moduli \cite{Acharya:2010zx}.
On the other hand, within this framework, we will see that both the overproduction and underproduction of dark matter abundance of the QCD axion (or ALPs) can be naturally accounted for through their mass mixing.

In the context of the axiverse, it is natural to consider the cosmological evolution of multiple axions, which has been extensively studied in recent years \cite{Hill:1988bu, Daido:2015cba, Cyncynates:2023esj, Li:2023uvt}.
When non-zero mass mixing between two axions is considered, a non-trivial cosmological evolution of the mass eigenvalues, known as axion level crossing, can occur during the QCD phase transition. 
Examples include the mass mixing between the QCD axion and an ALP \cite{Daido:2015cba}, between the QCD axion and a sterile axion \cite{Cyncynates:2023esj}, and between the $Z_{\mathcal N}$ axion \cite{Hook:2018jle, DiLuzio:2021pxd, DiLuzio:2021gos} and an ALP \cite{Li:2023uvt}, among others. 
Generally, level crossing can take place before the critical temperature of the QCD phase transition, $T_{\rm QCD}$. 
However, as highlighted in ref.~\cite{Li:2023uvt}, this is not always the case, as a second level crossing might occur at $T_{\rm QCD}$ in scenarios involving a $Z_{\mathcal N}$ axion --- a phenomenon referred to as double level crossings (see also ref.~\cite{Li:2024kdy, Li:2025imm} for other examples). 
The level crossing can induce an adiabatic transition in the axion energy density, analogous to the Mikheyev-Smirnov-Wolfenstein (MSW) effect \cite{Wolfenstein:1977ue, Mikheyev:1985zog, Mikheev:1986wj} observed in neutrino oscillations. 
It also carries profound cosmological implications, such as altering the axion relic density \cite{Ho:2018qur, Cyncynates:2021xzw, Murai:2023xjn, Li:2023xkn, Murai:2024nsp} and isocurvature perturbations \cite{Kitajima:2014xla, Daido:2015cba}, contributing to the composition of dark energy \cite{Muursepp:2024mbb, Muursepp:2024kcg}, influencing the formation of domain walls \cite{Daido:2015bva, Lee:2024toz, Li:2024psa}, and impacting gravitational waves and primordial black holes \cite{Li:2024psa, Cyncynates:2022wlq, Chen:2021wcf, Kitajima:2023cek, Lee:2024xjb, Li:2023det}, among other effects.

The examples mentioned above all consider mass mixing between two axions, with the selection of axions generally being relatively arbitrary.
In this work, we adopt a bottom-up perspective to investigate multi-axion mass mixing where the number of axions exceeds two.\footnote{See also $\rm e.g.$ refs.~\cite{Gendler:2023kjt, Chadha-Day:2023wub} for discussion on the topic of multi-axion mixing.}
Specifically, we will delve into several mixing scenarios, including the mixing of the QCD axion with one ALP (referred to as the $1+1$ mixing), the mixing of the QCD axion with two ALPs (referred to as the $1+2$ mixing), and the mixing of the QCD axion with multiple ALPs (referred to as the $1+N$ mixing, where $N>2$).
We will investigate what conditions an axion model must fulfill in order to exhibit maximal mixing.
Maximal mixing is defined as the mixing scenario where the axions experience the largest degree of effective mixing. 

The $1+1$ mixing scenario is a widely recognized framework that has garnered considerable attention and has been the subject of numerous previous studies.
However, it is still necessary to first clarify this scenario within the context of the string axiverse.
Primarily, this scenario focuses on the mass mixing that occurs between the QCD axion and an ALP.
More specifically, depending on the variation in the energy density of the QCD axion, it can be further categorized into two distinct scenarios: the light QCD axion scenario \cite{Daido:2015cba, Li:2023uvt} and the heavy QCD axion scenario \cite{Cyncynates:2023esj}.
Additionally, it is worth noting that scenarios involving the mixing between two QCD axions have also received significant attention within this framework \cite{Li:2024okl, Li:2024kdy}.
However, our proposed $1+2$ and $1+N$ mixing scenarios, which involve more than two axions within the string axiverse, have not been studied before. 

In the $1+2$ mixing scenario, we consider both isotropic and anisotropic cases. 
In the isotropic case, we consider the mixing of the QCD axion and two ALPs, with both ALPs having decay constants smaller than that of the QCD axion. 
On the other hand, in the anisotropic case, we can only consider the mixing of the QCD axion and one ALP, meaning that the $1+2$ mixing scenario reduces to the $1+1$ scenario. 
Specifically, the scenario where an ALP has a decay constant smaller than that of the QCD axion corresponds to the light QCD axion scenario, while the scenario with a larger decay constant corresponds to the heavy QCD axion scenario.
Additionally, we will briefly discuss the isotropic case where both ALPs have decay constants larger than that of the QCD axion.
Notice that in these mixing scenarios, the mass of the ALP is assumed to be constant and must be smaller than that of the QCD axion, and more precisely, than the zero-temperature QCD axion mass.

Subsequently, within the framework of the string axiverse, it is quite natural to extend the $1+2$ mixing scenario to the $1+N$ mixing scenario.
In this case, for effective mixing to occur, the decay constants of ALPs must be either smaller than or larger than that of the QCD axion. 
Additionally, the masses of ALPs cannot be equal; if two ALPs have the same mass, effective mixing will only take place once.
In the $1+N$ mixing scenario, we identify two key conditions for the occurrence of maximal mixing among multiple axion mass eigenstates:
\begin{itemize}
{\it 
\item The masses of all ALPs must be smaller than the zero-temperature mass of the QCD axion, and no two of the ALP masses can be equal; rather, there should be a distinct mass hierarchy among them.
\item The decay constants of all ALPs must simultaneously be either smaller than or larger than the decay constant of the QCD axion; otherwise, the maximal mixing scenario would be compromised.
}
\end{itemize}
Naturally, these conclusions are based on the analysis of the mixing scenario involving one QCD axion and multiple ALPs (all possessing constant masses) within the context of the type IIB string axiverse, but they should also be applicable to more generalized multi-axion mixing scenarios.
We also find that the transfer of axion energy density ultimately only occurs between the two axions with the closest masses, and this applies equally to both the $1+1$ and $1+2$ mixing scenarios.
In addition, the $1+N$ mixing scenario where all ALPs have decay constants larger than the QCD axion is also briefly discussed.
We also conduct a quantitative analysis of the maximal mixing and determine whether a given model exhibits this behavior.
 
Lastly, as mentioned earlier, axion mass mixing in the string axiverse exhibits rich phenomenology. 
We will briefly discuss its cosmological implications, including axion relic density, isocurvature fluctuations, dark energy, domain walls, gravitational waves, and primordial black holes. 
Notice that the string axiverse also possesses rich phenomenological aspects unrelated to axion mass mixing; see $\rm e.g.$ refs.~\cite{Arvanitaki:2010sy, Marsh:2011gr, Marsh:2013taa, Tashiro:2013yea, Yoshino:2014wwa, Kamionkowski:2014zda, Yoshino:2015nsa, Acharya:2015zfk, Emami:2016mrt, Karwal:2016vyq, Hardy:2016mns, Yoshida:2017cjl, Kitajima:2018zco, Agrawal:2019lkr, Calza:2021czr, Calza:2023rjt, Alexander:2023wgk, Gasparotto:2023psh, Martucci:2024trp, Das:2025eix}.

The rest of this paper is structured as follows.
In section~\ref{sec_axions_in_string_theory}, we provide a concise overview of axions in string theory.
In section~\ref{sec_IIB_string_axiverse}, we introduce the type IIB string axiverse. 
We first outline the concept of the LVS axiverse and then present several specific LVS axiverse models.
In section~\ref{sec_axion_mass_mixing}, we investigate axion mass mixing in the string axiverse, specifically, with the $1+1$, $1+2$, and $1+N$ mixing scenarios.
The cosmological implications of axion mass mixing are included in section~\ref{sec_cosmological_implications}.
Finally, the conclusion and outlook are given in section~\ref{sec_Conclusion}.

\section{Axions in string theory}%%%%%%%%%%%%%%%%%%%%%%%%%%
\label{sec_axions_in_string_theory}

In this section, we briefly review axions within the context of string theory \cite{Svrcek:2006yi}; see also ref.~\cite{Seo:2024zzs} for a recent overview.
There are two primary mechanisms in string theory for generating axions from $p$-form gauge fields $C_p$ upon compactification: model-independent and model-dependent axions.
The first mechanism involves dualizing the $2$-form gauge field in four non-compact spacetime dimensions, a process that is independent of the specific topological details of the internal manifold.
Conversely, model-dependent axions arise from $C_p$ components that have legs along the compact internal manifold; their existence and properties are thus dictated by the topological structure of the internal manifold.
 
\subsection{Model independent axions} 

Considering the Kalb-Ramond 2-form $B_2$ in heterotic string theory, the relevant effective action is given by
\begin{eqnarray}
-\dfrac{2\pi}{g_s^2 \ell_s^4}\int\dfrac{1}{2} H_3 \wedge \star_{10} H_3 - \dfrac{1}{8\pi g_s^2 \ell_s^6}\int {\rm Tr}(F\wedge \star_{10} F)\, ,
\end{eqnarray} 
where $\wedge$ is the wedge product, $\star$ is the Hodge star operation, $g_s$ and $\ell_s$ are the string coupling constant and length, respectively, and $H_3$ is the field strength of $B_2$ as
\begin{eqnarray} 
d H_3=\dfrac{1}{16\pi^2}\left[{\rm Tr}(R \wedge R)-{\rm Tr}(F \wedge F)\right]\, ,
\label{dH_3}
\end{eqnarray}
with the curvature 2-form $R$ and the gauge field strength $F$.
Notice that here $R \wedge R$ can be omitted and eq.~\eqref{dH_3} can be imposed on the action by introducing the 6-form $B_6$.
Then, by integrating out $H_3$ and dualizing to $B_6$, we can obtain
\begin{eqnarray}  
-\dfrac{g_s^2 \ell_s^4}{2\pi}\int\dfrac{1}{2} H_7 \wedge \star_{10} H_7 - \dfrac{1}{8\pi g_s^2 \ell_s^6}\int {\rm Tr}(F\wedge \star_{10} F) + \dfrac{1}{16\pi^2}\int B_6 \wedge {\rm Tr}(F \wedge F) \, ,
\label{action_H_7}
\end{eqnarray} 
where $H_7\equiv dB_6=2\pi/(g_s^2 \ell_s^4)\star_{10} H_3$.
Here, we consider the heterotic string compactified to four dimensions on a compact internal manifold $K_6$.
After the KK reduction, the first and second terms in eq.~\eqref{action_H_7} can be written as
\begin{eqnarray} 
-\int_{X_4}\dfrac{1}{2} f^2 d\theta\wedge\star_4d\theta+\dfrac{1}{8\pi^2}\sum_a \int_{X_4}\dfrac{1}{2}\theta {\rm Tr}(F^{(a)} \wedge F^{(a)}) \, ,
\end{eqnarray} 
where $X_4$ represents the four-dimensional non-compact manifold and the axion decay constant $f$ is given by
\begin{eqnarray}
f^2=\dfrac{g_s^2 \ell_s^4}{2\pi {\rm Vol}(K_6)}=\dfrac{g_s^2}{2\pi \ell_s^2 \mathcal{V}}\, ,
\end{eqnarray} 
with the volume of the internal manifold ${\rm Vol}(K_6)=\ell_s^6 \mathcal{V}$.
 
\subsection{Model dependent axions}  
 
Furthermore, string theory encompasses various $p$-form gauge fields $C_p$ where $p$ is not restricted to $p=2$. 
Upon compactification on $p$-cycles, these fields can produce four-dimensional axions. 
The KK expansion of $C_p$ is given by
\begin{eqnarray}
C_p(x,y)=\sum_{\alpha=1}^{b_p(K_6)} \dfrac{\theta^\alpha (x)}{2\pi} \omega_{p,\alpha}(y)\, ,
\end{eqnarray} 
where $\omega_{p,\alpha}$ ($\alpha=1$, $\cdots$, $b_p(K_6)$) represent the harmonic forms.
Using the inner product of two harmonic forms 
\begin{eqnarray}
\langle \omega_{p,\alpha}, \omega_{p,\beta} \rangle {\rm Vol}(K_6)=\int_{K_6} \omega_{p,\alpha}\wedge \star_6 \omega_{p,\beta} \, ,
\end{eqnarray} 
we have the kinetic term for $C_p$ in the string frame 
\begin{eqnarray}
-\dfrac{g_s^2}{2k_{10}^2}\int\dfrac{1}{2}d C_p\wedge \star d C_p=-\int_{X_4}\dfrac{1}{2} f_{\alpha\beta}^2 d \theta^\alpha \wedge \star_4 d \theta^\beta\, ,
\end{eqnarray}  
where $k_{10}$ is the ten-dimensional gravitational coupling and the axion decay constant $f_{\alpha\beta}$ is given by 
\begin{eqnarray}
f_{\alpha\beta}^2=\dfrac{g_s^2}{8\pi^2}M_{\rm Pl}^2 \langle \omega_{p,\alpha}, \omega_{p,\beta} \rangle\, ,
\end{eqnarray} 
with the four-dimensional reduced Planck mass $M_{\rm Pl}$.
When the compact internal manifold $K_6$ is a CY three-fold, the resulting four-dimensional EFT displays $\mathcal{N}=1$ supersymmetry in type I and heterotic string theories, whereas it exhibits $\mathcal{N}=2$ supersymmetry in type II string theory.
In these cases, the axion and the Kähler modulus reside in the same supersymmetry multiplet and therefore can be considered as the pseudoscalar and scalar components of a complex scalar field, respectively.
Notice that some axions are projected out of the low-energy spectrum.

\section{Review of the type IIB string axiverse}%%%%%%%%%%%%%%%%%%%%%%%%%%
\label{sec_IIB_string_axiverse}

In this section, we introduce the type IIB string axiverse \cite{Cicoli:2012sz, Cicoli:2011qg, Cicoli:2012cy}.
To begin with, we will introduce the concept of the LVS axiverse, providing a foundational understanding.
Subsequently, we will present several specific LVS axiverse models.

\subsection{The type IIB LVS axiverse}

In type IIB flux compactifications on CY, the SM is localized on a stack of D3 or D7 branes. 
The pseudo-scalar axion-like fields $c^a$ ($a=1, \cdots ,h_-^{1,1}$) and $c_\alpha$ ($\alpha=1, \cdots ,h_+^{1,1}$) can arise as KK zero modes of the Ramond-Ramond (RR) antisymmetric tensor fields $C_2$ and $C_4$.  
Here we consider the case with the Hodge number $h_-^{1,1}=0$, $\rm i.e.$, all the moduli corresponding to $h_-^{1,1}$ are projected out by the orientifold involution, and thus $h^{1,1}=h_+^{1,1}$.
The Kähler moduli can be described by
\begin{eqnarray}
T_\alpha=\tau_\alpha+{\rm i}\, c_\alpha\, ,
\end{eqnarray} 
where $\tau_\alpha$ are the divisor volumes; see appendix~\ref{app_divisor_volumes}.
To obtain the axion decay constant, we first need to perform an orthogonal transformation for the effective Lagrangian (see also appendix~\ref{app_axion_couplings}), which results in the kinetic term \cite{Cicoli:2012sz}
\begin{eqnarray}
\mathcal{L} \supset -\dfrac{1}{8}\lambda_\alpha\partial_\mu c'_\alpha \partial^\mu c'_\alpha \, ,
\end{eqnarray} 
where $\lambda_\alpha$ are the eigenvalues of $\mathcal{K}_{\alpha\beta}$.
Subsequently, we need to define new fields for canonical normalization
\begin{eqnarray}
a_\alpha\equiv\dfrac{1}{2}\sqrt{\lambda_\alpha} c'_\alpha\, ,
\end{eqnarray} 
and finally, we obtain the decay constants for canonically normalized axions\footnote{Notice that, in the context of efficient mass mixing, this quantity may not accurately represent the true field range of the axion.} \cite{Cicoli:2012sz}
\begin{eqnarray}
f_\alpha\equiv\dfrac{\sqrt{\lambda_\alpha}M_{\rm Pl}}{4\pi}\, ,
\end{eqnarray} 
with the periodicity $a_\alpha=a_\alpha+2\pi f_\alpha$.

Considering the simplest Swiss-cheese CY manifold, we have the volume \cite{Denef:2004dm}
\begin{eqnarray}
\mathcal{V}=\dfrac{1}{9\sqrt{2}}\left(\tau_b^{3/2}-\tau_s^{3/2}\right)\, ,
\end{eqnarray} 
where $\tau_b$ and $\tau_s$ are the large and small divisor volumes, respectively.
In the limit $\tau_b\gg \tau_s$, the decay constants of the corresponding axions are given by \cite{Cicoli:2012sz}
\begin{eqnarray}
f_{a_b}=\dfrac{\sqrt{3}M_{\rm Pl}}{4\pi\tau_b} \simeq\dfrac{M_{\rm Pl}}{4\pi \mathcal{V}^{2/3}}\, , \quad f_{a_s}=\dfrac{1}{\sqrt{6}\left(2\tau_s\right)^{1/4}} \dfrac{M_{\rm Pl}}{4\pi \sqrt{\mathcal{V}}} \simeq\dfrac{M_s}{\sqrt{4\pi}\tau_s^{1/4}}\, ,
\end{eqnarray} 
where $M_s$ is the string scale.
Now, considering fibred CY manifolds with a del Pezzo divisor, we have the volume \cite{Cicoli:2011it}
\begin{eqnarray}
\mathcal{V}=t_b t_f^2+t_s^3\, ,
\end{eqnarray} 
where $\tau_b=2t_b t_f$, $\tau_f=t_f^2$, and $\tau_s=3t_s^2$ are the divisor volumes, respectively.
In the limit $t_s \tau_f\gg \tau_s^{3/2}$, the decay constants of the corresponding axions are given by \cite{Cicoli:2012sz}
\begin{eqnarray}
f_{a_f}=\dfrac{M_{\rm Pl}}{4\pi\tau_f}\, , \quad f_{a_b}=\dfrac{\sqrt{2}M_{\rm Pl}}{4\pi\tau_b}\, , \quad f_{a_s}=\dfrac{1}{\sqrt{2}\left(3\tau_s\right)^{1/4}} \dfrac{M_{\rm Pl}}{4\pi \sqrt{\mathcal{V}}} \simeq\dfrac{M_s}{\sqrt{4\pi}\tau_s^{1/4}}\, .
\end{eqnarray}
Furthermore, there are two scenarios in which the CY can have either an isotropic or anisotropic shape.
In the isotropic limit $t_b\sim \sqrt{\tau_f}\sim \mathcal{V}^{1/3}$, we have 
\begin{eqnarray}
f_{a_f}\simeq f_{a_b} \simeq\dfrac{M_{\rm Pl}}{4\pi \mathcal{V}^{2/3}}\, , \quad f_{a_s}\simeq\dfrac{M_s}{\sqrt{4\pi}\tau_s^{1/4}}\, .
\end{eqnarray} 
While in the anisotropic limit $t_b\sim \mathcal{V} \gg \sqrt{\tau_f}\sim \sqrt{\tau_s}$, we have
\begin{eqnarray}
f_{a_f}\simeq\dfrac{M_{\rm Pl}}{4\pi\tau_s}\, , \quad f_{a_b}\simeq\dfrac{M_{\rm Pl}}{4\pi \tau_b}\, , \quad f_{a_s}\simeq\dfrac{M_s}{\sqrt{4\pi}\tau_s^{1/4}}\, .
\end{eqnarray}
Here we present only the axion decay constants. 
Another crucial factor is the coupling strength between axion and gauge bosons, especially considering the topology of the CY three-fold and the choice of brane set-up and fluxes, which determines whether it can behave as the QCD axion with an anomaly coefficient of order unity.
More details about the type IIB string axiverse can be found in appendix~\ref{app_type_IIB_string_axiverse}.

\subsection{Specific LVS axiverse models}
 
In this subsection, we present a range of specific LVS axiverse models sourced from refs.~\cite{Cicoli:2012sz, Cicoli:2011qg, Cicoli:2012cy}. 
Notice that these models provide a theoretical basis for the investigations in subsequent sections.
Therefore, it is necessary to have a preliminary understanding of them.
Here, we show three distinct types of models: firstly, a model containing one QCD axion plus one ALP; secondly, a model featuring one QCD axion plus two ALPs; and lastly, models that encompass one QCD axion plus multiple ALPs.

\subsubsection{One QCD axion plus one ALP}
\label{sec_IIB_string_axiverse_1+1}
 
We first introduce a model containing one QCD axion plus one ALP from refs.~\cite{Cicoli:2012sz, Cicoli:2011qg}. 
This is an anisotropic $\rm SU(3)\times SU(2)$ model with $d = 1$ D-term equation leading to $n_{\rm ax} = 2$ light axions: a fibre axion as the QCD axion, and a base axion as the light ALP.
Notice that another isotropic model exists, but it has only the QCD axion.

The internal space of this type IIB model is an orientifold of a K3 fibred CY three-fold, characterized by specific Hodge numbers $h^{1,1}=h_+^{1,1}=4$, realized as a hypersurface in a toric variety. 
The key divisors for modeling include the K3 fiber $D_1$, two intersecting rigid four-cycles $D_4$ and $D_5$, and a del Pezzo divisor $D_7$.
The Kähler form is given by \cite{Cicoli:2011qg}
\begin{eqnarray}
J=t_1 \hat{D}_1+t_4 \hat{D}_4+t_5 \hat{D}_5+t_7 \hat{D}_7\, ,
\end{eqnarray}
and the overall volume is 
\begin{eqnarray}
\mathcal{V}=\dfrac{1}{2}t_{\rm base}\tau_1+t_4 t_5^2-\dfrac{1}{3}t_5^3-\dfrac{1}{3}\tau_7^{3/2}\, ,
\end{eqnarray}
where $t_{\rm base}=2(t_1-t_5)$.
The $\rm SU(3)\times SU(2)$ model is accomplished by wrapping 3 D7-branes around $D_4$ and 1 D7-brane around the K3 fiber $D_1$.
The anomalous $\rm U(1)$ acquires mass through the Stückelberg mechanism by eaten up a closed string axion.
The gauge flux generates a moduli-dependent Fayet-Iliopoulos (FI) term, resulting in the fixation of one Kähler modulus at $\tau_4=3(\tau_1-\tau_5)-\tau_7$.
The three remaining Kähler moduli are $\tau_1$, $\tau_5$, and $\tau_7$.
By integrating out the D-term fixed modulus, we can determine the masses of light axions using an effective volume \cite{Cicoli:2011qg}
\begin{eqnarray}
\mathcal{V}=\dfrac{1}{3}\left(\sqrt{\tau_s}\tau_b-\tau_7^{3/2}\right)\, ,
\end{eqnarray}
where $\tau_s\equiv \tau_1-\tau_5$, and $\tau_b\equiv (10\tau_1-\tau_5)/2$.
The interplay between the non-perturbative effect and the leading order $\alpha'$ correction to the Kähler potential fixes $\sqrt{\tau_s}\tau_b$ and $\tau_7$ according to the standard LVS approach.
The non-perturbative effects on $D_7$ generate a potential for the axion $c_7$, thereby giving $a_7$ a heavy mass and rendering it not the QCD axion. 
Then the two remaining massless axions are $a_s$ and $a_b$, which remain massless due to the $\alpha'$ and $g_s$ corrections.
In the anisotropic limit $\tau_b\gg \tau_s\sim \tau_7$, the decay constants of the corresponding QCD axion and light ALP are given by \cite{Cicoli:2012sz}
\begin{eqnarray}
{\rm QCD~axion}\!&:&~f_{a_s}\simeq \dfrac{M_{\rm Pl}}{4\pi \tau_s}\simeq 1\times 10^{16}\, \rm GeV\, ,\\
{\rm ALP}\!&:&~f_{a_b}\simeq \dfrac{M_{\rm Pl}}{4\pi \tau_b}\simeq 5\times 10^{3}\, \rm GeV\, .
\end{eqnarray}
We can observe that the large decay constant of the axion $a_s$ does not fall within the classic QCD axion window as depicted in eq.~\eqref{axion_classical_window}. 
However, its abundance can be diluted here through the late out-of-equilibrium decay of $\tau_7$ and $a_7$ \cite{Acharya:2010zx}. 
Furthermore, in this context, the abundance of the QCD axion can be naturally and significantly suppressed in the scenario involving mass mixing with the ALP, which we will discuss in the subsequent sections.

\subsubsection{One QCD axion plus two ALPs}
\label{sec_IIB_string_axiverse_1+2}

Here we introduce a model containing one QCD axion plus two ALPs from refs.~\cite{Cicoli:2012sz, Cicoli:2012cy}. 
In contrast to the previous discussion, the QCD axion in this context possesses a decay constant with the intermediate-scale $\sim\mathcal{O}(10^{10}) \, \rm GeV$.

The internal space of this model is an orientifold of a K3 or a $T^4$ fibred CY three-fold, characterized by specific Hodge numbers $h^{1,1}=h_+^{1,1}=5$, realized as a hypersurface in a toric variety. 
The key divisors for modeling include the K3 or $T^4$ fiber $D_1$, a divisor $D_2$ that controls the base of the fibration, a del Pezzo four-cycle $D_3$, and two rigid divisors $D_4$ and $D_5$.
The Kähler form is given by \cite{Cicoli:2012cy}
\begin{eqnarray}
J=t_1 \hat{D}_1+t_2 \hat{D}_2-t_3 \hat{D}_3-t_4 \hat{D}_4-t_5 \hat{D}_5\, ,
\end{eqnarray}
and the overall volume is 
\begin{eqnarray}
\mathcal{V}=\alpha\left[\sqrt{\tau_1}\tau_2-\gamma_3\tau_3^{3/2}-\gamma_5\tau_5^{3/2}-\gamma_4\left(\tau_4-x\tau_5\right)^{3/2}\right]\, ,
\end{eqnarray}
where $\alpha$, $\gamma_i$, and $x$ are constants of order one.
The moduli-dependent FI terms result in a fixed combination of $\tau_4$ and $\tau_5$ at $\tau_5=\lambda\hat{\tau}_4$, where $\lambda\equiv(x\gamma_4/\gamma_5)^2$, and $\hat{\tau}_4\equiv\tau_4-x \tau_5$.
Then the four remaining Kähler moduli are $\tau_1$, $\tau_2$, $\tau_3$, and $\hat{\tau}_4$.
After the D-term stabilization, we have the eﬀective volume  
\begin{eqnarray}
\mathcal{V}=\alpha\left(\sqrt{\tau_1}\tau_2-\gamma_3\tau_3^{3/2}-\lambda_4\hat{\tau}_4^{3/2}\right)\, ,
\end{eqnarray}
where $\lambda_4\equiv\gamma_5\lambda^{3/2}+\gamma_4$.
The non-perturbative effects on $D_3$ generate a potential for the axion $c_3$, thereby giving $a_3$ a heavy mass and rendering it not the QCD axion.
The $\alpha'$ and $g_s$ corrections result in the three remaining massless axions being $a_1$, $a_2$, and $a_4$. 
The decay constant of the axion $a_4$ is given by \cite{Cicoli:2012sz}
\begin{eqnarray}
{\rm QCD~axion}\!:~f_{a_4}\simeq \dfrac{M_s}{\sqrt{4\pi}}\simeq 1\times 10^{10}\, \rm GeV\, ,
\end{eqnarray} 
which is a perfect QCD axion candidate with an intermediate-scale decay constant.
The axions $a_1$ and $a_2$ both belong to the light ALPs, and they can be discussed within the framework of two scenarios: isotropic and anisotropic.
In the isotropic case $\tau_1\sim\tau_2~\mathcal{V}^{2/3}$, the decay constants of $a_1$ and $a_2$ are given by
\begin{eqnarray}
{\rm ALPs}\!:~f_{a_1}\simeq f_{a_2}\simeq \dfrac{M_{\rm Pl}}{4\pi \mathcal{V}^{2/3}}\simeq 1\times 10^{8}\, \rm GeV\, .
\end{eqnarray} 
While in the anisotropic case $\tau_2\gg\tau_1\sim\hat{\tau}_4$, the decay constants are given by
\begin{eqnarray}
{\rm ALPs}\!:~f_{a_1}\simeq \dfrac{M_{\rm Pl}}{4\pi \tau_1}\simeq 1\times 10^{16}\, {\rm GeV}\, , \quad f_{a_2}\simeq \dfrac{M_{\rm Pl}}{4\pi \tau_2}\simeq 5\times 10^{3}\, {\rm GeV}\, .   
\label{f_a_anisotropic}
\end{eqnarray}
Notice that there is another anisotropic scenario, where the SM D7-stack wraps around the fiber divisor $D_1$, featuring $a_2$ as the ALP and $a_1$ taking the role of the QCD axion.
In this context, with the SM residing on $D_1$, $a_4$ transforms into a heavy axion due to non-perturbative effects that are no longer canceled by the chiral intersection with SM matter fields.\footnote{We are grateful to Michele Cicoli for clarifying this point.}
In the subsequent discussions, we will focus only on the anisotropic scenario involving two ALPs presented in eq.~\eqref{f_a_anisotropic}.

\subsubsection{One QCD axion plus multiple ALPs}
\label{sec_IIB_string_axiverse_1+N} 

In the previous two subsections, we introduced models containing one QCD axion plus one ALP and two ALPs, respectively. 
It is natural to extend this scenario to a model incorporating one QCD axion plus multiple ALPs. 
Here, we will provide only a brief overview of such an extension from ref.~\cite{Cicoli:2012sz}.
In this case, we require extra local cycles intersecting with visible branes, without imposing additional D-term conditions. 
Given $n$ intersecting local cycles, of which $d$ are fixed by D-terms, this leaves $n-d$ moduli frozen by loop corrections. 
Consequently, this results in $n-d$ light axions with intermediate-scale decay constants.
For our purposes, in the subsequent sections, we will consider scenarios where the QCD axion and multiple ALPs have comparable decay constants, specifically with $f\sim \mathcal{O}(10^{10})\, \rm GeV$.
In addition, there exist several explicit and globally consistent CY models, which feature the QCD axion and a small number of light ALPs \cite{Cicoli:2012vw, Cicoli:2013mpa, Cicoli:2013cha}.
Recently, an LVS axiverse model featuring the QCD axion and multiple ultra-light ALPs was proposed in ref.~\cite{Broeckel:2021dpz}, where the number of axions can reach $\mathcal{O}(100)$. 
This approach ensures the attainment of a perfect QCD axion decay constant at the TeV scale. 
In this model, the required hierarchies in axion masses and decay constants can be naturally realized; see ref.~\cite{Broeckel:2021dpz} for further details.

\section{Axion mass mixing in the string axiverse}%%%%%%%%%%%%%%%%%%%%%%%%%%
\label{sec_axion_mass_mixing}

In this section, we investigate axion mass mixing in the string axiverse and the conditions an axion model must fulfill in order to exhibit maximal mixing, providing a quantitative analysis.
Based on the above discussions, we will consider the following mixing scenarios here: $1+1$, $1+2$, and $1+N$ ($N>2$), where the first number before the plus sign indicates the number of the QCD axion, and the numbers after the plus sign represent the number of ALPs. 
The decay constants of axions in the mixing scenarios, which originate from the string axiverse, are listed in table~\ref{tab_1}. 
Notice that the subscript ``0$"$ used hereinafter represents the QCD axion $a_0$, whereas the subscript denoting a positive real number stands for the ALP $A_i$. 
The subscript convention should not be confused with the subscripts employed in the previous sections.
 
%%%%%%%%%%%%%%%%%%%%%%%%%%%%%%%%
\begin{table}[t]
\centering
\begin{tabular}{lcccc}
\hline\hline
Scenarios   &    $f_{a_0}$ ($\rm GeV$)   &    $f_{A_1}$ ($\rm GeV$)   &  $f_{A_2}$ ($\rm GeV$)  &   $f_{A_N}$ ($\rm GeV$) \\
\hline
$1+1$ &  $1\times10^{16}$     &  $5\times10^{3}$   & --- &   --- \\
$1+2$ (isotropic) &  $1\times10^{10}$     &  $1\times10^{8}$   & $1\times10^{8}$ &   --- \\
$1+2$ (anisotropic) &  $1\times10^{10}$     &  $5\times10^{3}$   & $1\times10^{16}$ &   --- \\
$1+N$&   $\sim1\times10^{10}$     &  $\sim1\times10^{10}$   & $\sim1\times10^{10}$ &   $\sim1\times10^{10}$  \\
 \hline\hline
\end{tabular}
\caption{The decay constants of axions in the mixing scenarios.
Notice that, in the maximal $1+N$ mixing scenario, the decay constants of all ALPs should simultaneously be either smaller than or larger than the decay constant of the QCD axion.}
\label{tab_1}
\end{table}
%%%%%%%%%%%%%%%%

\subsection{The $1+1$ mixing scenario} 
\label{sec_axion_mass_mixing_1+1} 
 
Let us first discuss the so-called $1+1$ mixing scenario. 
Notably, this is a well-studied scenario that has garnered considerable attention and has been extensively studied in numerous pieces of previous literature.
Primarily, this scenario focuses on the mass mixing that occurs between the QCD axion and an ALP, see $\rm e.g.$ refs.~\cite{Daido:2015cba, Cyncynates:2023esj, Li:2023uvt}. 
Furthermore, it is worth noting that scenarios involving the mixing between two QCD axions have also been given considerable consideration within the framework of this discussion; see also refs.~\cite{Li:2024okl, Li:2024kdy}.
 
In a class of string axiverse models similar to that discussed in section~\ref{sec_IIB_string_axiverse_1+1} with two axions, in this context, the hierarchical decay constants of the QCD axion ($a_0$) and ALP ($A_1$) are given by
\begin{eqnarray}
f_{a_0}=1\times 10^{16}\, {\rm GeV}\, , \quad f_{A_1}=5\times 10^{3}\, \rm GeV\, .
\end{eqnarray}
The low-energy effective Lagrangian that describes this $1+1$ mixing scenario can be formulated as follows
\begin{eqnarray}
\begin{aligned}
\mathcal{L}&\supset\dfrac{1}{2} f_{a_0}^2 \left(\partial\theta_0\right)^2 + \dfrac{1}{2} f_{A_1}^2 \left(\partial\Theta_1\right)^2\\
&-m_{a_0}^2 f_{a_0}^2\left[1-\cos\left(n_{00}\theta_0+n_{01}\Theta_1+\delta_0\right)\right]\\
&-m_{A_1}^2 f_{A_1}^2\left[1-\cos\left(n_{10}\theta_0+n_{11}\Theta_1+\delta_1\right)\right]\, ,
\label{Lagrangian_1+1}
\end{aligned}
\end{eqnarray}
where $\theta_0=\phi_0/f_{a_0}$ and $\Theta_1=\varphi_1/f_{A_1}$ represent the QCD axion and ALP angles, respectively, $\phi_0$ and $\varphi_1$ are the QCD axion and ALP fields, $m_{a_0}$ and $m_{A_1}$ are the corresponding axion masses, $n_{ij}$ are the domain wall numbers, and $\delta_i$ are the constant phases.
Depending on the choice of domain wall numbers, this mixing scenario can be further categorized into the light QCD axion scenario and the heavy QCD axion scenario, which we will discuss in section~\ref{sec_axion_mass_mixing_1+2_anisotropic}.
In this and subsequent discussions, we assume the constant phases to be zero, which can be seen as equivalent to demanding an independent solution for the strong CP problem.
This assumption does not affect the evolution of axion fields during the mixing process and thus is justified within the context of our discussions.
Notice that the single-field ALP is regarded as the most straightforward scenario, characterized by a constant mass $m_{A_1}$, whereas the QCD axion mass $m_{a_0}$ is temperature-dependent; see also appendix~\ref{app_axion_relic_density_without_mixing}.
In addition, in scenarios where there is strong mass mixing in the potential, the QCD axion will exhibit the PQ quality problem and cease to be a true QCD axion.

For our purpose, here the domain wall numbers $n_{ij}$ should be taken as
\begin{eqnarray}
n_{00}=1\, , \quad n_{01}=0\, , \quad n_{10}=1\, , \quad n_{11}=1\, .
\end{eqnarray}
To facilitate a more intuitive understanding, it is essential for us to define a matrix representing the domain wall numbers\footnote{In this matrix representation, the elements are arranged according to their subscripts, with the entry in the $i$-th row and $j$-th column representing $n_{ij}$; this arrangement applies to subsequent matrices of domain wall numbers as well.}
\begin{eqnarray}
\mathfrak{n}_{2\times2}=
\left(
\begin{array}{cc}
1  & ~ 0\\
1  & ~ 1
\end{array}
\right)\, .
\end{eqnarray}
Furthermore, the axion masses should follow the relation
\begin{eqnarray}
m_{A_1}<m_{a_0,0}\, ,
\end{eqnarray}
where $m_{a_0,0}$ is the zero-temperature QCD axion mass.
After substituting the mixing potential $V_{\rm mix}$ derived from eq.~\eqref{Lagrangian_1+1} into the equations of motion (EOMs), we obtain the following differential equations governing the evolution of axion fields
\begin{eqnarray}
\ddot\phi_0+3H\dot\phi_0+\dfrac{\partial V_{\rm mix}(\phi_0, \varphi_1)}{\partial \phi_0}=0\, ,
\end{eqnarray}
\begin{eqnarray}
\ddot\varphi_1+3H\dot\varphi_1+\dfrac{\partial V_{\rm mix}(\phi_0, \varphi_1)}{\partial \varphi_1}=0\, ,
\end{eqnarray}
where the dot notation denotes differentiation with respect to the physical time $t$ (one dot for the first derivative and two dots for the second derivative), $\partial_{\phi_0} V_{\rm mix}$ and $\partial_{\varphi_1} V_{\rm mix}$ represent the partial derivatives of the mixing potential $V_{\rm mix}$ with respect to the axion fields, respectively, and $H(T)$ is the Hubble parameter. Specifically, for the fields $\phi_0$ and $\varphi_1$, the EOMs expand to
\begin{eqnarray}
\ddot\phi_0+3H\dot\phi_0+ m_{a_0}^2 f_{a_0}\sin\left(\dfrac{\phi_0}{f_{a_0}}\right)+ \dfrac{m_{A_1}^2 f_{A_1}^2}{f_{a_0}}\sin\left(\dfrac{\phi_0}{f_{a_0}}+\dfrac{\varphi_1}{f_{A_1}}\right)=0 \, , 
\end{eqnarray}
\begin{eqnarray}
\ddot\varphi_1+3H\dot\varphi_1+ m_{A_1}^2 f_{A_1}\sin\left(\dfrac{\phi_0}{f_{a_0}}+\dfrac{\varphi_1}{f_{A_1}}\right)=0\, , 
\end{eqnarray}
respectively. 
Assuming that the oscillation amplitudes of the axion fields are significantly smaller than their respective decay constants, we can approximate the system and derive the mass mixing matrix within this model
\begin{eqnarray}
\mathbf{M}^2=
\left(\begin{array}{cc}
m_{a_0}^2+\dfrac{m_{A_1}^2 f_{A_1}^2}{f_{a_0}^2}  & ~ \dfrac{m_{A_1}^2 f_{A_1}}{f_{a_0}}\\
\dfrac{m_{A_1}^2 f_{A_1}}{f_{a_0}} & ~ m_{A_1}^2
\end{array}\right)\, .
\end{eqnarray}
By diagonalizing this matrix, we can identify the heavy and light mass eigenstates $a_h$ and $a_l$, respectively, through the transformation
\begin{eqnarray}
\left(\begin{array}{c}
a_h \\
a_l 
\end{array}\right)
=\left(\begin{array}{cc}
\cos \alpha & \quad \sin \alpha \\
-\sin \alpha  & \quad   \cos \alpha
\end{array}\right)
\left(\begin{array}{c}
\phi_0 \\
\varphi_1 
\end{array}\right)\, ,
\end{eqnarray} 
where $\alpha$ is the mass mixing angle
\begin{eqnarray}
\cos^2\alpha=\dfrac{1}{2}\left(1+\dfrac{m_{A_1}^2-m_{a_0}^2-\dfrac{m_{A_1}^2 f_{A_1}^2}{f_{a_0}^2}}{m_h^2-m_l^2}\right) \, ,
\end{eqnarray} 
and $m_h$ and $m_l$ (with the assumption that $m_h > m_l$) are the corresponding mass eigenvalues associated with the heavy and light mass eigenstates, respectively,
\begin{eqnarray}
\begin{aligned}
m_{h,l}^2&=\dfrac{1}{2}\left[m_{a_0}^2+m_{A_1}^2+\dfrac{m_{A_1}^2 f_{A_1}^2}{f_{a_0}^2}\right]\\
&\pm\dfrac{1}{2 f_{a_0}^2}\bigg[-4 m_{a_0}^2 m_{A_1}^2 f_{a_0}^4+\bigg(\left(m_{a_0}^2+ m_{A_1}^2\right)f_{a_0}^2+m_{A_1}^2 f_{A_1}^2\bigg)^2\bigg]^{1/2}\, .
\label{mass_eigenvalues}
\end{aligned}
\end{eqnarray}  

%%%%%%%%%%%%%%%%%%%%%%%%%%%%%%%%% 
\begin{figure}[t]%%%%%%%%%%%%%%%%%%%%%%%%%% 
\centering
\includegraphics[width=0.70\textwidth]{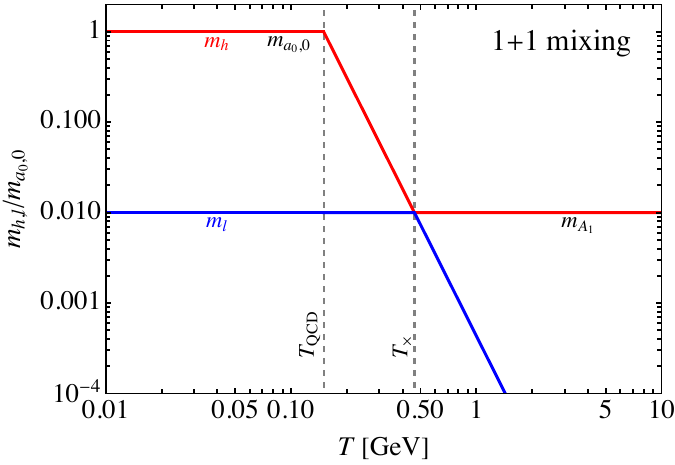}
\caption{The illustration of the $1+1$ mixing scenario.
The solid lines represent the normalized temperature-dependent axion mass eigenvalues $m_h(T)$ and $m_l(T)$, respectively.
The vertical gray lines (from right to left) indicate the temperatures $T_\times$ and $T_{\rm QCD}$, respectively.
The masses $m_{a_0,0}$ and $m_{A_1}$ are also labeled.
The model parameters can be found in tables~\ref{tab_1} and \ref{tab_2}.}
\label{fig_me_1+1}
\end{figure}

In figure~\ref{fig_me_1+1}, we provide a detailed depiction of the mass eigenvalues $m_{h,l}$. 
For comparison with the subsequent mixing scenarios, here we present the normalized eigenvalues $m_{h,l}/m_{a_0,0}$.
It is noteworthy that a level crossing phenomenon may occur when the difference between the squared masses, denoted as $m_h^2-m_l^2$, attains its minimal value. 
To determine the precise temperature at which this level crossing occurs within the mass mixing, we can solve the differential equation
\begin{eqnarray}
\dfrac{d\left(m_h^2(T)-m_l^2(T)\right)}{dT}\bigg|_{T>T_{\rm QCD}}=0\, .
\label{differential_equation}
\end{eqnarray}
Upon solving this equation, we derive the level crossing temperature, $T_\times$, which is expressed as follows\footnote{A straightforward derivation is provided in appendix~\ref{app_level_crossing_temperature}.}
\begin{eqnarray}
T_\times= T_{\rm QCD} \left(\dfrac{m_{a_0,0}^2 f_{a_0}^2}{m_{A_1}^2 \left(f_{a_0}^2-f_{A_1}^2\right)}\right)^{\frac{1}{2b}} \, . 
\label{Tx_1+1}  
\end{eqnarray}
From this formula, we can also find that the condition $f_{a_0}^2-f_{A_1}^2>0$ must be satisfied first. 
In this scenario, the evolution of axions is as follows. 
At high cosmic temperatures, the mass eigenvalues $m_l$ and $m_h$ are associated with the QCD axion $a_0$ and the ALP $A_1$, respectively. 
As the temperature gradually decreases towards $T_\times$, a level crossing phenomenon occurs between the QCD axion and the ALP. 
Consequently, this level crossing results in $m_l$ switching its association from the QCD axion to the ALP, while $m_h$ correspondingly shifts its alignment from the ALP to the QCD axion. 
This sequential pattern of mixing continues as the temperature steadily declines.  
If the adiabatic condition is met, the axion energy density will undergo an adiabatic transfer at the level crossing \cite{Ho:2018qur, Murai:2024nsp}.
Note also that this scenario can effectively suppress the QCD axion abundance and is therefore referred to as the light QCD axion scenario; see refs.~\cite{Daido:2015cba, Li:2023uvt} for more details. 

%%%%%%%%%%%%%%%%%%%%%%%%%%%%%%%%
\begin{table}[t]
\centering
\begin{tabular}{lcccccc}
\hline\hline
Scenarios            & $r_{A_1}$ & $r_{A_2}$ & $r_{A_3}$ & $\cdots$ & $r_{A_{N-1}}$ & $r_{A_{N}}$\\
\hline
$1+1$                      & 0.01        & ---        &  ---    & & ---   &---\\
$1+2$ (isotropic)     &  0.0025   & 0.005   &  ---    & & ---   &---\\
$1+2$ (anisotropic) &  ---           & 0.01     &  ---    & & ---   &---\\
$1+N$                     &  0.001     &  0.003  &0.009  & $\cdots$ & 0.1  & 0.3 \\
 \hline\hline
\end{tabular}
\caption{The typical masses of ALPs in the mixing scenarios. Here we introduce the mass ratio $r_{A_i}$, defined as $r_{A_i}\equiv m_{A_i}/m_{a_0,0}$, where $m_{a_0,0}$ is the zero-temperature QCD axion mass.}
\label{tab_2}
\end{table}
%%%%%%%%%%%%%%%%

\subsection{The $1+2$ mixing scenario}
\label{sec_axion_mass_mixing_1+2}

In the subsequent subsections, we will present novel investigations into the scenario where the number of ALPs in the mass mixing exceeds one.

\subsubsection{Isotropic case}
\label{sec_axion_mass_mixing_1+2_isotropic}

In a class of string axiverse models similar to that discussed in section~\ref{sec_IIB_string_axiverse_1+2} with three axions, the decay constants of the QCD axion ($a_0$) and ALPs ($A_1$, $A_2$) are given by
\begin{eqnarray}
f_{a_0}=1\times 10^{10}\, {\rm GeV}\, , \quad f_{A_1}=1\times 10^{8}\, {\rm GeV}\, , \quad f_{A_2}=1\times 10^{8}\, {\rm GeV}\, .
\end{eqnarray}
Notice that the decay constants of both ALPs here are smaller than that of the QCD axion.
The low-energy effective Lagrangian that describes this $1+2$ mixing scenario can be formulated as follows
\begin{eqnarray}
\begin{aligned}
\mathcal{L}&\supset\dfrac{1}{2} f_{a_0}^2 \left(\partial\theta_0\right)^2 + \dfrac{1}{2} f_{A_1}^2 \left(\partial\Theta_1\right)^2 + \dfrac{1}{2} f_{A_2}^2 \left(\partial\Theta_2\right)^2\\
&-m_{a_0}^2 f_{a_0}^2\left[1-\cos\left(n_{00}\theta+n_{01}\Theta_1+n_{02}\Theta_2+\delta_0\right)\right]\\
&-m_{A_1}^2 f_{A_1}^2\left[1-\cos\left(n_{10}\theta+n_{11}\Theta_1+n_{12}\Theta_2+\delta_1\right)\right]\\
&-m_{A_2}^2 f_{A_2}^2\left[1-\cos\left(n_{20}\theta+n_{21}\Theta_1+n_{22}\Theta_2+\delta_2\right)\right]\, ,
\label{Lagrangian_1+2}
\end{aligned}
\end{eqnarray}
where $\Theta_1=\varphi_1/f_{A_1}$ and $\Theta_2=\varphi_2/f_{A_2}$ represent the ALP angles.
Given that our focus is on the most straightforward scenario of the single-field ALP characterized by a constant mass, we find that, in circumstances involving axion mass mixing, the mixing process is confined to at most a single instance. 
Specifically, the QCD axion will mix with each individual ALP no more than once.
This understanding is based on the assumption that we are dealing with the simplest form of ALP fields.

For our purpose, here the matrix of domain wall numbers should be taken as 
\begin{eqnarray}
\mathfrak{n}_{3\times3}=
\left(
\begin{array}{ccc}
1  & ~ 0 & ~ 0\\
1  & ~ 1 & ~ 0\\
1  & ~ 0 & ~ 1
\end{array}
\right)\, .
\label{dw_matrix_1+2_light}
\end{eqnarray}
Furthermore, the axion masses should follow the relation
\begin{eqnarray}
m_{A_1}<m_{a_0,0}\, , \quad m_{A_2}<m_{a_0,0}\, ,\quad m_{A_1}\neq m_{A_2}\, .
\label{relation_ma_1+2}
\end{eqnarray}
Notice that, in order to more effectively demonstrate the intricacies of the mixing process, we hereby make the assumption that the masses of these two ALPs are unequal. 
In practical scenarios, if their masses were indeed identical, the mixing dynamics would be significantly simplified, leading to a mixing scenario with just one interaction; specifically, the QCD axion would mix solely with one of the ALPs, without the complexity of multiple mass-dependent interactions.
Then the EOMs expand to
\begin{eqnarray}
\begin{aligned}
\ddot\phi_0+3H\dot\phi_0+ m_{a_0}^2 f_{a_0}\sin\left(\dfrac{\phi_0}{f_{a_0}}\right)+ \dfrac{m_{A_1}^2 f_{A_1}^2}{f_{a_0}}\sin\left(\dfrac{\phi_0}{f_{a_0}}+\dfrac{\varphi_1}{f_{A_1}}\right)\\
+\dfrac{m_{A_2}^2 f_{A_2}^2}{f_{a_0}}\sin\left(\dfrac{\phi_0}{f_{a_0}}+\dfrac{\varphi_2}{f_{A_2}}\right)=0 \, , 
\end{aligned}
\end{eqnarray}
\begin{eqnarray}
\ddot\varphi_1+3H\dot\varphi_1+ m_{A_1}^2 f_{A_1}\sin\left(\dfrac{\phi_0}{f_{a_0}}+\dfrac{\varphi_1}{f_{A_1}}\right)=0\, , 
\end{eqnarray}
\begin{eqnarray}
\ddot\varphi_2+3H\dot\varphi_2+ m_{A_2}^2 f_{A_2}\sin\left(\dfrac{\phi_0}{f_{a_0}}+\dfrac{\varphi_2}{f_{A_2}}\right)=0\, , 
\end{eqnarray}
respectively.
The mass mixing matrix is given by
\begin{eqnarray}
\mathbf{M}^2=
\left(\begin{array}{ccc}
m_{a_0}^2+\dfrac{m_{A_1}^2 f_{A_1}^2+m_{A_2}^2 f_{A_2}^2}{f_{a_0}^2} & ~ \dfrac{m_{A_1}^2 f_{A_1}}{f_{a_0}} & ~ \dfrac{m_{A_2}^2 f_{A_2}}{f_{a_0}}\\
\dfrac{m_{A_1}^2 f_{A_1}}{f_{a_0}}   & ~ m_{A_1}^2 & ~ 0\\
\dfrac{m_{A_2}^2 f_{A_2}}{f_{a_0}}   & ~ 0 & ~ m_{A_2}^2
\end{array}\right)\, .
\end{eqnarray}
As discussed earlier, since we are considering the mixing of the QCD axion with each ALP separately, the aforementioned mass mixing matrix can be expressed as the following effective mixing matrices
\begin{eqnarray}
\mathbf{M}_1^2=
\left(\begin{array}{cc}
m_{a_0}^2+\dfrac{m_{A_1}^2 f_{A_1}^2}{f_{a_0}^2}  & ~ \dfrac{m_{A_1}^2 f_{A_1}}{f_{a_0}}\\
\dfrac{m_{A_1}^2 f_{A_1}}{f_{a_0}} & ~ m_{A_1}^2
\end{array}\right)\, ,\quad 
\mathbf{M}_2^2=
\left(\begin{array}{cc}
m_{a_0}^2+\dfrac{m_{A_2}^2 f_{A_2}^2}{f_{a_0}^2}  & ~ \dfrac{m_{A_2}^2 f_{A_2}}{f_{a_0}}\\
\dfrac{m_{A_2}^2 f_{A_2}}{f_{a_0}} & ~ m_{A_2}^2
\end{array}\right)\, ,~
\end{eqnarray} 
with the corresponding mass eigenvalues, respectively,  
\begin{eqnarray}
\begin{aligned}
m_{h_1,l_1}^2&=\dfrac{1}{2}\left[m_{a_0}^2+m_{A_1}^2+\dfrac{m_{A_1}^2 f_{A_1}^2}{f_{a_0}^2}\right]\\
&\pm\dfrac{1}{2 f_{a_0}^2}\bigg[-4 m_{a_0}^2 m_{A_1}^2 f_{a_0}^4+\bigg(\left(m_{a_0}^2+ m_{A_1}^2\right)f_{a_0}^2+m_{A_1}^2 f_{A_1}^2\bigg)^2\bigg]^{1/2}\, ,
\end{aligned}
\end{eqnarray} 
\begin{eqnarray}
\begin{aligned}
m_{h_2,l_2}^2&=\dfrac{1}{2}\left[m_{a_0}^2+m_{A_2}^2+\dfrac{m_{A_2}^2 f_{A_2}^2}{f_{a_0}^2}\right]\\
&\pm\dfrac{1}{2 f_{a_0}^2}\bigg[-4 m_{a_0}^2 m_{A_2}^2 f_{a_0}^4+\bigg(\left(m_{a_0}^2+ m_{A_2}^2\right)f_{a_0}^2+m_{A_2}^2 f_{A_2}^2\bigg)^2\bigg]^{1/2}\, .
\end{aligned}
\end{eqnarray} 

%%%%%%%%%%%%%%%%%%%%%%%%%%%%%%%%% 
\begin{figure}[t]%%%%%%%%%%%%%%%%%%%%%%%%%% 
\centering
\includegraphics[width=0.70\textwidth]{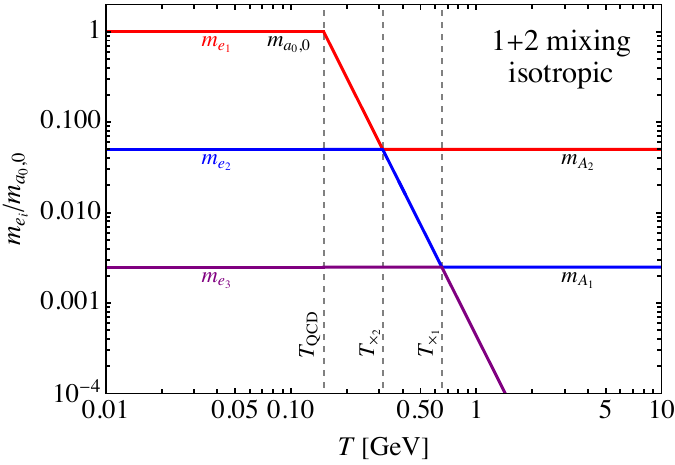}
\caption{The illustration of the $1+2$ mixing scenario (isotropic).
The solid lines represent the normalized temperature-dependent axion mass eigenvalues $m_{e_1}(T)$, $m_{e_2}(T)$, and $m_{e_3}(T)$, respectively.
The vertical gray lines (from right to left) indicate the temperatures $T_{\times_1}$, $T_{\times_2}$, and $T_{\rm QCD}$, respectively.
The masses $m_{a_0,0}$, $m_{A_1}$, and $m_{A_2}$ are also labeled.
The model parameters can be found in tables~\ref{tab_1} and \ref{tab_2}.}
\label{fig_me_1+2_isotropic}
\end{figure}

Based on eq.~\eqref{relation_ma_1+2}, if we consider a case where the mass of the ALP $A_1$ is smaller than that of $A_2$,
\begin{eqnarray}
m_{A_1}<m_{A_2}\, ,
\end{eqnarray} 
the mass eigenvalues $m_{e}$ can be expressed as
\begin{eqnarray}
m_{e_1}&=&m_{h_2}\, ,\\
m_{e_2}&\simeq&
\begin{cases}
m_{l_2}\, ,  &  T \le T_{\times}^{(m)}\\
m_{h_1}\, , &  T > T_{\times}^{(m)}
\end{cases}  \\
m_{e_3}&=&m_{l_1}\, ,
\end{eqnarray}
with
\begin{eqnarray}
T_{\times_2}<T_{\times}^{(m)}<T_{\times_1}\, ,
\end{eqnarray}
where $T_{\times}^{(m)}$ represents an intermediate temperature between the level crossing temperatures $T_{\times_1}$ and $T_{\times_2}$.
In figure~\ref{fig_me_1+2_isotropic}, we show the mass eigenvalues in this scenario.
The red, blue, and purple solid lines correspond to the normalized eigenvalues $m_{e_1}$, $m_{e_2}$, and $m_{e_3}$, respectively.
Here the level crossing temperatures can be obtained by solving the differential equations
\begin{eqnarray}
\dfrac{d\left(m_{e_2}^2(T)-m_{e_3}^2(T)\right)}{dT}\bigg|_{T>T_{\times}^{(m)}}=0\, ,
\label{differential_equations_Tx_1+2_1}
\end{eqnarray}
\begin{eqnarray}
\dfrac{d\left(m_{e_1}^2(T)-m_{e_2}^2(T)\right)}{dT}\bigg|_{T_{\rm QCD}<T<T_{\times}^{(m)}}=0\, ,
\label{differential_equations_Tx_1+2_2}
\end{eqnarray}
then we have 
\begin{eqnarray}
T_{\times_1}= T_{\rm QCD} \left(\dfrac{m_{a_0,0}^2 f_{a_0}^2}{m_{A_1}^2 \left(f_{a_0}^2-f_{A_1}^2\right)}\right)^{\frac{1}{2b}} \, ,
\label{Tx_1+2_1}  
\end{eqnarray}
\begin{eqnarray}
T_{\times_2}= T_{\rm QCD} \left(\dfrac{m_{a_0,0}^2 f_{a_0}^2}{m_{A_2}^2 \left(f_{a_0}^2-f_{A_2}^2\right)}\right)^{\frac{1}{2b}} \, .
\label{Tx_1+2_2}  
\end{eqnarray}
We can find that $T_{\times_1}$ is identical to the level crossing temperature presented in eq.~\eqref{Tx_1+1}.
Note also that $T_{\times}^{(m)}$ is an estimated value, used to control a specific temperature range, and its exact value is not required.
Next, we analyze the evolution of multiple axions in this scenario.
At high cosmic temperatures, the mass eigenvalues $m_{e_3}$, $m_{e_2}$, and $m_{e_1}$ are respectively associated with the QCD axion $a_0$, the ALP $A_1$, and the ALP $A_2$. 
As the temperature decreases to $T_{\times_1}$, a level crossing occurs between the QCD axion and $A_1$, resulting in $m_{e_3}$ now corresponding to $A_1$, while $m_{e_2}$ shifting its correspondence to the QCD axion.
As the temperature further descends to $T_{\times_2}$, another level crossing takes place, this time between the QCD axion and the ALP $A_2$. 
Consequently, $m_{e_2}$ now corresponds to $A_2$, and $m_{e_1}$ takes on the correspondence of the QCD axion. 
This sequence of mixings persists as the temperature continues to decline.

A notable feature of our findings is that $m_{e_2}$ undergoes two transformations: initially corresponding to the ALP $A_1$, then transitioning to the QCD axion $a_0$, and ultimately to the ALP $A_2$ --- a sequence that has not been previously observed in axion mass mixing.
In contrast, the double level crossing scenario in ref.~\cite{Li:2023uvt} exhibits a transition where, despite two sequential transformations, the axion ultimately reverts to its initial state.
Conversely, in the scenario we are discussing, the mass eigenstate initially aligns with one axion and ultimately with another distinct axion, highlighting a profound and essential distinction between these two scenarios. 
More importantly, our focus here is specifically on the mixing involving three distinct axions.

Additionally, we highlight an important aspect of the axion energy density transition.
In this $1+2$ mixing scenario, we observe three transfer processes: first, the transfer from the QCD axion $a_0$ to the ALP $A_1$; second, the transfer from $A_1$ to $a_0$, followed by the transfer from $a_0$ to the ALP $A_2$; and third, the transfer from $A_2$ to the QCD axion. 
This differs from the mutual transition between the QCD axion and ALP discussed in section~\ref{sec_axion_mass_mixing_1+1}.

\subsubsection{Anisotropic case}
\label{sec_axion_mass_mixing_1+2_anisotropic}

In a class of string axiverse models similar to that discussed in section~\ref{sec_IIB_string_axiverse_1+2}, the decay constants of the QCD axion ($a_0$) and the two ALPs ($A_1$, $A_2$) are given by
\begin{eqnarray}
f_{a_0}=1\times 10^{10}\, {\rm GeV}\, , \quad f_{A_1}=5\times 10^{3}\, {\rm GeV}\, , \quad f_{A_2}=1\times 10^{16}\, {\rm GeV} \, .
\end{eqnarray}
In contrast to section~\ref{sec_axion_mass_mixing_1+2_isotropic}, the decay constants of the two ALPs presented here exhibit a variation, with one being smaller and the other larger than that of the QCD axion.
Therefore, in this scenario, we consider the mixing involving either the QCD axion and the ALP $A_1$, or the QCD axion and the ALP $A_2$, rather than considering both simultaneously as was done in section~\ref{sec_axion_mass_mixing_1+2_isotropic}.
That is to say, in this case, the $1+2$ mixing scenario effectively reduces to the $1+1$ mixing scenario.

\paragraph{Light QCD axion scenario}

Here, we first consider the mixing of the QCD axion and the ALP $A_1$, with their decay constants
\begin{eqnarray}
f_{a_0}=1\times 10^{10}\, {\rm GeV}\, , \quad f_{A_1}=5\times 10^{3}\, {\rm GeV}\, .
\end{eqnarray}
One can find that the mass mixing in this scenario is identical to the $1+1$ mixing scenario discussed in section~\ref{sec_axion_mass_mixing_1+1}; therefore, we will not delve further into it here.
This mixing is referred to as the light QCD axion scenario \cite{Daido:2015cba, Li:2023uvt}. 
Next, we will proceed to discuss an alternative, which is the heavy QCD axion scenario \cite{Cyncynates:2023esj}.

\paragraph{Heavy QCD axion scenario}

\begin{figure}[t]%%%%%%%%%%%%%%%%%%%%%%%%%% 
\centering
\includegraphics[width=0.70\textwidth]{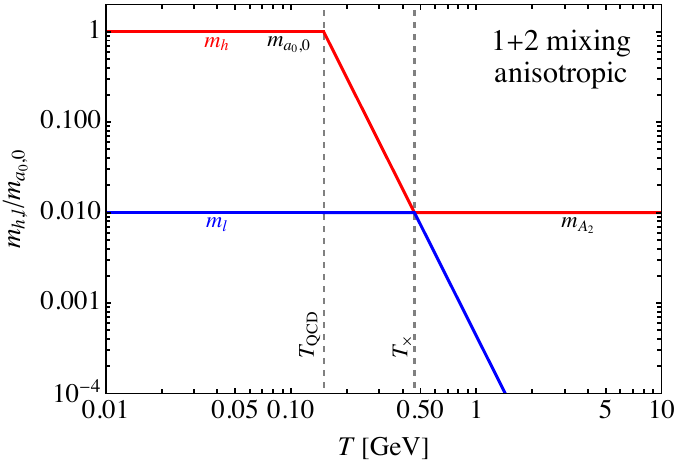}
\caption{The illustration of the $1+2$ mixing scenario (anisotropic).
The solid lines represent the normalized temperature-dependent axion mass eigenvalues $m_h(T)$ and $m_l(T)$, respectively.
The model parameters can be found in tables~\ref{tab_1} and \ref{tab_2}.
Notice that this is actually a $1+1$ mixing scenario.}
\label{fig_me_1+2_anisotropic}
\end{figure}

Then we discuss the mixing of the QCD axion and the ALP $A_2$ with their decay constants
\begin{eqnarray}
f_{a_0}=1\times 10^{10}\, {\rm GeV}\, , \quad f_{A_2}=1\times 10^{16}\, {\rm GeV}\, .
\end{eqnarray}
The low-energy effective Lagrangian that describes this heavy QCD axion scenario can be formulated as follows
\begin{eqnarray}
\begin{aligned}
\mathcal{L}&\supset\dfrac{1}{2} f_{a_0}^2 \left(\partial\theta_0\right)^2 + \dfrac{1}{2} f_{A_2}^2 \left(\partial\Theta_2\right)^2\\
&-m_{a_0}^2 f_{a_0}^2\left[1-\cos\left(n_{00}\theta_0+n_{01}\Theta_2+\delta_0\right)\right]\\
&-m_{A_2}^2 f_{A_2}^2\left[1-\cos\left(n_{20}\theta_0+n_{21}\Theta_2+\delta_2\right)\right]\, .
\label{Lagrangian_1+2_heavy}
\end{aligned}
\end{eqnarray}
For our purpose, here the matrix of domain wall numbers should be taken as
\begin{eqnarray}
\mathfrak{n}_{2\times2}=
\left(
\begin{array}{cc}
1  & ~ 1\\
0  & ~ 1
\end{array}
\right)\, ,
\end{eqnarray}
and the axion masses should follow the relation
\begin{eqnarray}
m_{A_2}<m_{a_0,0}\, .
\end{eqnarray}
The EOMs expand to
\begin{eqnarray}
\ddot\phi_0+3H\dot\phi_0+ m_{a_0}^2 f_{a_0}\sin\left(\dfrac{\phi_0}{f_{a_0}}+\dfrac{\varphi_2}{f_{A_2}}\right)=0 \, , 
\end{eqnarray}
\begin{eqnarray}
\ddot\varphi_2+3H\dot\varphi_2+ m_{A_2}^2 f_{A_2}\sin\left(\dfrac{\varphi_2}{f_{A_2}}\right)+\dfrac{m_{a_0}^2 f_{a_0}^2}{f_{A_2}}\sin\left(\dfrac{\phi_0}{f_{a_0}}+\dfrac{\varphi_2}{f_{A_2}}\right)=0\, , 
\end{eqnarray}
respectively. 
The mass mixing matrix in this case is given by
\begin{eqnarray}
\mathbf{M}^2=
\left(\begin{array}{cc}
m_{a_0}^2  & ~ \dfrac{m_{a_0}^2 f_{a_0}}{f_{A_2}}\\
\dfrac{m_{a_0}^2 f_{a_0}}{f_{A_2}} & ~ m_{A_2}^2+\dfrac{m_{a_0}^2 f_{a_0}^2}{f_{A_2}^2}
\end{array}\right)\, ,
\end{eqnarray}
with the mass eigenvalues
\begin{eqnarray}
\begin{aligned}
m_{h,l}^2&=\dfrac{1}{2}\left[m_{a_0}^2+m_{A_2}^2+\dfrac{m_{a_0}^2 f_{a_0}^2}{f_{A_2}^2}\right]\\
&\pm\dfrac{1}{2 f_{A_2}^2}\bigg[-4 m_{a_0}^2 m_{A_2}^2 f_{A_2}^4+\bigg(\left(m_{a_0}^2+ m_{A_2}^2\right)f_{A_2}^2+m_{a_0}^2 f_{a_0}^2\bigg)^2\bigg]^{1/2}\, .
\end{aligned}
\end{eqnarray}  
The level crossing temperature in this case is given by
\begin{eqnarray}
T_\times= T_{\rm QCD} \left(\dfrac{m_{a_0,0}^2 \left(f_{a_0}^2+f_{A_2}^2\right)^2}{m_{A_2}^2 f_{A_2}^2\left(f_{A_2}^2-f_{a_0}^2\right)}\right)^{\frac{1}{2b}} \, . 
\label{Tx_1+2}  
\end{eqnarray}
Given that the fundamental nature of this particular scenario is a $1+1$ mixing scenario, and the axion evolution process resembles that discussed in section~\ref{sec_axion_mass_mixing_1+1}, we will not delve into further details here.
In figure~\ref{fig_me_1+2_anisotropic}, we also show the mass eigenvalues in this scenario.
It is worth noting that, unlike the light QCD axion scenario shown in section~\ref{sec_axion_mass_mixing_1+1}, this scenario effectively enhances the QCD axion abundance and is therefore referred to as the heavy QCD axion scenario. 
For a comprehensive understanding of this scenario and its applications, please refer to refs.~\cite{Cyncynates:2023esj, Li:2024jko}.

%%%%%%%%%%%%%%%%%%%%%%%%%%%%%%%%% 

Recalling the mixing scenario presented in section~\ref{sec_axion_mass_mixing_1+2_isotropic}, it is readily apparent that it corresponds to the light QCD axion scenario. 
If we consider the case where the decay constants of ALPs in section~\ref{sec_axion_mass_mixing_1+2_isotropic} are all larger than that of the QCD axion --- that is, the heavy QCD axion scenario --- then the matrix of domain wall numbers should be taken as 
\begin{eqnarray}
\mathfrak{n}_{3\times3}=
\left(
\begin{array}{ccc}
1  & ~ 1 & ~ 1\\
0  & ~ 1 & ~ 0\\
0  & ~ 0 & ~ 1
\end{array}
\right)\, .
\label{dw_matrix_1+2_heavy}
\end{eqnarray}
Here, we will not discuss further details about this scenario, but will only present the mass mixing matrix
\begin{eqnarray}
\mathbf{M}^2=
\left(\begin{array}{ccc}
m_{a_0}^2 & ~ \dfrac{m_{a_0}^2 f_{a_0}}{f_{A_1}} & ~ \dfrac{m_{a_0}^2 f_{a_0}}{f_{A_2}}\\
\dfrac{m_{a_0}^2 f_{a_0}}{f_{A_1}}   & ~ m_{A_1}^2+\dfrac{m_{a_0}^2 f_{a_0}^2}{f_{A_1}^2} & ~ \dfrac{m_{a_0}^2 f_{a_0}^2}{f_{A_1} f_{A_2}}\\
\dfrac{m_{a_0}^2 f_{a_0}}{f_{A_2}}   & ~ \dfrac{m_{a_0}^2 f_{a_0}^2}{f_{A_1} f_{A_2}} & ~ m_{A_2}^2+\dfrac{m_{a_0}^2 f_{a_0}^2}{f_{A_2}^2}
\end{array}\right)\, ,
\end{eqnarray}
with the effective mixing matrices
\begin{eqnarray}
\mathbf{M}_1^2=
\left(\begin{array}{cc}
m_{a_0}^2  & ~ \dfrac{m_{a_0}^2 f_{a_0}}{f_{A_1}}\\
\dfrac{m_{a_0}^2 f_{a_0}}{f_{A_1}} & ~ m_{A_1}^2+\dfrac{m_{a_0}^2 f_{a_0}^2}{f_{A_1}^2}
\end{array}\right)\, ,\quad 
\mathbf{M}_2^2=
\left(\begin{array}{cc}
m_{a_0}^2  & ~ \dfrac{m_{a_0}^2 f_{a_0}}{f_{A_2}}\\
\dfrac{m_{a_0}^2 f_{a_0}}{f_{A_2}} & ~ m_{A_2}^2+\dfrac{m_{a_0}^2 f_{a_0}^2}{f_{A_2}^2}
\end{array}\right)\, .
\end{eqnarray}

\subsection{The $1+N$ mixing scenario}
\label{sec_axion_mass_mixing_1+N}

In this subsection, we generalize the previously discussed $1+2$ mixing scenario to the $1+N$ ($N>2$) mixing scenario.
 
In a class of string axiverse models similar to that discussed in section~\ref{sec_IIB_string_axiverse_1+N} with multiple axions, the decay constants of the QCD axion ($a_0$) and the ALPs ($A_i$), in which $A_i$ are numbered from 1 to $N$, are approximately given by
\begin{eqnarray}
f_{a_0}\sim f_{A_i}\sim 1\times10^{10}\, \rm GeV\, .
\end{eqnarray}
Here we first discuss the scenario where the decay constants of ALPs are all smaller than that of the QCD axion,
\begin{eqnarray}
f_{A_i}<f_{a_0},\, \forall i\, .
\label{condition_fa_N+1}
\end{eqnarray}
The low-energy effective Lagrangian that describes this $1+N$ mixing scenario can be formulated as follows
\begin{eqnarray}
\begin{aligned}
\mathcal{L}&\supset\dfrac{1}{2} f_{a_0}^2 \left(\partial\theta_0\right)^2 + \dfrac{1}{2} \sum_{i=1}^N f_{A_i}^2 \left(\partial\Theta_i\right)^2\\
&-m_{a_0}^2 f_{a_0}^2\left[1-\cos\left(n_{00}\theta_0+\sum_{j=1}^N n_{0j}\Theta_j+\delta_0\right)\right]\\
&-\sum_{i=1}^N m_{A_i}^2 f_{A_i}^2\left[1-\cos\left(n_{i0}\theta_0+\sum_{j=1}^N n_{ij}\Theta_j+\delta_i\right)\right]\, ,
\label{Lagrangian_1+N}
\end{aligned}
\end{eqnarray}
where $\Theta_i=\varphi_i/f_{A_i}$ are the ALP angles, $\varphi_i$ represent the ALP fields, $f_{A_i}$ and $m_{A_i}$ are the ALP decay constants and masses, respectively.
 
For our purpose, here the matrix of domain wall numbers should be taken as
\begin{eqnarray}
\mathfrak{n}_{(N+1)\times(N+1)}=
\left(
\begin{array}{cccccc}
1  & ~ 0 & ~ 0 & ~ 0 &~\cdots & ~ 0\\
1  & ~ 1 & ~ 0 & ~ 0 &~\cdots & ~ 0\\
1  & ~ 0 & ~ 1 & ~ 0 &~\cdots & ~ 0\\
1  & ~ 0 & ~ 0 & ~ 1 &~\cdots & ~ 0\\
\vdots  & ~\vdots & ~\vdots &~\vdots & ~\ddots & ~ \vdots\\
1  & ~ 0 & ~ 0 & ~ 0 & ~ \cdots & ~ 1
\end{array}
\right)\, ,
\end{eqnarray}
and the axion masses should follow the relation
\begin{eqnarray}
m_{A_i}<m_{a_0,0},\, \forall i\, , \quad m_{A_i}\neq m_{A_j},\, \forall i\neq j\, .
\label{relation_ma_1+N}
\end{eqnarray}
Then the EOMs expand to
\begin{eqnarray}
\ddot\phi_0+3H\dot\phi_0+ m_{a_0}^2 f_{a_0}\sin\left(\dfrac{\phi_0}{f_{a_0}}\right)+ \sum_{i=1}^N\dfrac{m_{A_i}^2 f_{A_i}^2}{f_{a_0}}\sin\left(\dfrac{\phi_0}{f_{a_0}}+\dfrac{\varphi_i}{f_{A_i}}\right)=0 \, , 
\end{eqnarray}
\begin{eqnarray}
\ddot\varphi_1+3H\dot\varphi_1+ m_{A_1}^2 f_{A_1}\sin\left(\dfrac{\phi_0}{f_{a_0}}+\dfrac{\varphi_1}{f_{A_1}}\right)=0\, , 
\end{eqnarray}
\begin{eqnarray}
\ddot\varphi_2+3H\dot\varphi_2+ m_{A_2}^2 f_{A_2}\sin\left(\dfrac{\phi_0}{f_{a_0}}+\dfrac{\varphi_2}{f_{A_2}}\right)=0\, , 
\end{eqnarray}
$\hspace{7cm}\vdots$
\begin{eqnarray}
\ddot\varphi_N+3H\dot\varphi_N+ m_{A_N}^2 f_{A_N}\sin\left(\dfrac{\phi_0}{f_{a_0}}+\dfrac{\varphi_N}{f_{A_N}}\right)=0\, , 
\end{eqnarray}
respectively, with the mass mixing matrix
\begin{eqnarray}
\mathbf{M}^2=
\left(
\begin{array}{ccccc}
m_{a_0}^2+\dfrac{1}{f_{a_0}^2} \displaystyle\sum_{i=1}^{N} m_{A_i}^2 f_{A_i}^2& ~ \dfrac{m_{A_1}^2 f_{A_1}}{f_{a_0}} & ~ \dfrac{m_{A_2}^2 f_{A_2}}{f_{a_0}} & ~\cdots &~ \dfrac{m_{A_N}^2 f_{A_N}}{f_{a_0}}\\
\dfrac{m_{A_1}^2 f_{A_1}}{f_{a_0}}   & ~ m_{A_1}^2 & ~ 0 & ~\cdots & ~0\\
\dfrac{m_{A_2}^2 f_{A_2}}{f_{a_0}}   & ~ 0 & ~ m_{A_2}^2 & ~\cdots & ~0\\
\vdots  & ~\vdots & ~\vdots  & ~\ddots & ~\vdots\\
\dfrac{m_{A_N}^2 f_{A_N}}{f_{a_0}}   & ~ 0 & ~ 0 & ~\cdots & ~ m_{A_N}^2
\end{array}
\right)\, .
\end{eqnarray}
As discussed earlier, the effective mixing matrices can be separately expressed as
\begin{eqnarray}
\mathbf{M}_i^2=
\left(\begin{array}{cc}
m_{a_0}^2+\dfrac{m_{A_i}^2 f_{A_i}^2}{f_{a_0}^2}  & ~ \dfrac{m_{A_i}^2 f_{A_i}}{f_{a_0}}\\
\dfrac{m_{A_i}^2 f_{A_i}}{f_{a_0}} & ~ m_{A_i}^2
\end{array}\right)\, ,
\end{eqnarray} 
with the corresponding mass eigenvalues
\begin{eqnarray}
\begin{aligned}
m_{h_i,l_i}^2&=\dfrac{1}{2}\left[m_{a_0}^2+m_{A_i}^2+\dfrac{m_{A_i}^2 f_{A_i}^2}{f_{a_0}^2}\right]\\
&\pm\dfrac{1}{2 f_{a_0}^2}\bigg[-4 m_{a_0}^2 m_{A_i}^2 f_{a_0}^4+\bigg(\left(m_{a_0}^2+ m_{A_i}^2\right)f_{a_0}^2+m_{A_i}^2 f_{A_i}^2\bigg)^2\bigg]^{1/2}\, .
\end{aligned}
\end{eqnarray} 

\begin{figure}[t]%%%%%%%%%%%%%%%%%%%%%%%%%% 
\centering
\includegraphics[width=0.70\textwidth]{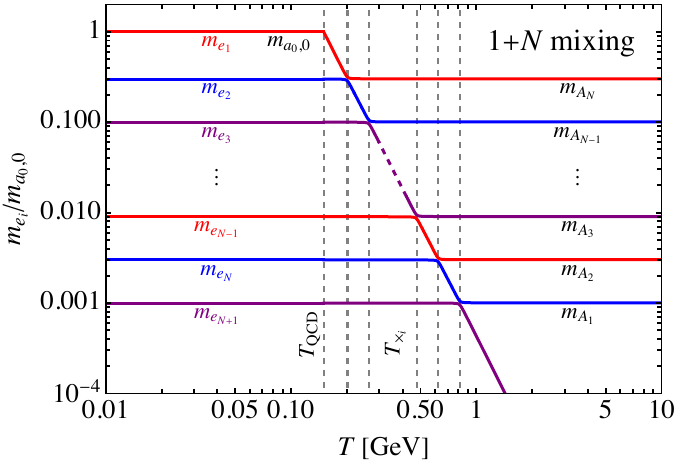}
\caption{The illustration of the $1+N$ mixing scenario.
The solid lines represent the normalized temperature-dependent axion mass eigenvalues $m_{e_i}(T)$, respectively.
The vertical gray lines (from right to left) indicate the temperatures $T_{\times_i}$ and $T_{\rm QCD}$, respectively.
The masses $m_{a_0,0}$ and $m_{A_i}$ are also labeled.
Here we set $f_{a_0}=1\times10^{10}\, \rm GeV$ and $f_{A_i}=1\times10^{9}\, \rm GeV$.
The other model parameters can be found in table~\ref{tab_2}.}
\label{fig_mhl_1+N}
\end{figure}

If we consider a case where the mass of $A_i$ is smaller than that of $A_{i+1}$,
\begin{eqnarray}
m_{A_1}<m_{A_2}<\cdots<m_{A_N}\, ,
\label{m_A_i_1+N}
\end{eqnarray} 
the mass eigenvalues $m_{e_i}$ can be expressed as
\begin{eqnarray}
m_{e_1}&=&m_{h_N}\, ,\\
m_{e_2}&\simeq&
\begin{cases}
m_{l_N}\, ,  &  T \le T_{\times_{N-1}}^{(m)}\\
m_{h_{N-1}}\, , &  T > T_{\times_{N-1}}^{(m)}
\end{cases}  \\
m_{e_3}&\simeq&
\begin{cases}
m_{l_{N-1}}\, ,  &  T \le T_{\times_{N-2}}^{(m)}\\
m_{h_{N-2}}\, , &  T > T_{\times_{N-2}}^{(m)}
\end{cases}  \\
&\vdots\nonumber\\
m_{e_N}&\simeq&
\begin{cases}
m_{l_2}\, ,  &  T \le T_{\times_1}^{(m)}\\
m_{h_1}\, , &  T > T_{\times_1}^{(m)}
\end{cases}  \\
m_{e_{N+1}}&=&m_{l_1}\, ,
\end{eqnarray}
with
\begin{eqnarray}
T_{\times_{i+1}}<T_{\times_i}^{(m)}<T_{\times_i}\, ,
\end{eqnarray}
where $T_{\times_i}^{(m)}$ represent the corresponding intermediate temperatures.
In this case, the level crossing temperatures $T_{\times_i}$ can be obtained by solving the differential equations
\begin{eqnarray}
\dfrac{d\left(m_{e_N}^2(T)-m_{e_{N+1}}^2(T)\right)}{dT}\bigg|_{T>T_{\times_1}^{(m)}}&=&0\, , \quad i=1\\
\dfrac{d\left(m_{e_{N-i+1}}^2(T)-m_{e_{N-i+2}}^2(T)\right)}{dT}\bigg|_{T_{\times_i}^{(m)}<T<T_{\times_{i-1}}^{(m)}}&=&0\, , \quad 1< i< N\\
\dfrac{d\left(m_{e_1}^2(T)-m_{e_2}^2(T)\right)}{dT}\bigg|_{T_{\rm QCD}<T<T_{\times_{N-1}}^{(m)}}&=&0\, , \quad i=N
\end{eqnarray}
then we have 
\begin{eqnarray}
T_{\times_i}= T_{\rm QCD} \left(\dfrac{m_{a_0,0}^2 f_{a_0}^2}{m_{A_i}^2 \left(f_{a_0}^2-f_{A_i}^2\right)}\right)^{\frac{1}{2b}} \, .
\label{Tx_1+N}  
\end{eqnarray}
In figure~\ref{fig_mhl_1+N}, we show the mass eigenvalues in this scenario; the solid lines correspond to the normalized $m_{e_i}$. 
We can observe that the evolution process of axions here is similar to that described in section~\ref{sec_axion_mass_mixing_1+2_isotropic}, except for the significantly greater number of ALPs involved here, leading to multiple level crossings. 
Notice that in this figure, the decay constants of all ALPs have been assumed to be identical. 
Furthermore, these decay constants are not significantly smaller than that of the QCD axion, in contrast to the scenarios depicted in figures~\ref{fig_me_1+1} and \ref{fig_me_1+2_isotropic}. 
Consequently, the level crossing phenomenon demonstrated here is more pronounced.

%%%%%%%%%%%%%%%%%%%%%%%%%%%%%%%%% 
 
Here we summarize and delineate two key conditions that are essential for the occurrence of maximal mixing among multiple axion mass eigenstates.
As defined in the introduction, maximal mixing is characterized as the mixing scenario where the axions experience the largest degree of effective mixing.  
{\it Firstly}, the masses of all ALPs must be strictly smaller than the zero-temperature mass of the QCD axion. 
In addition, it is crucial to note that there should be no equivalence in the masses of any two ALPs; rather, they should exhibit a distinct hierarchy among themselves. 
This hierarchical distribution ensures that no two ALPs can interfere with each other's mixing dynamics, thereby facilitating maximal mixing.
{\it Secondly}, the decay constants of all ALPs play a crucial role in determining the extent of maximal mixing. 
Specifically, the decay constants of all ALPs must uniformly be either smaller than or larger than the decay constant of the QCD axion.\footnote{Logically, this condition can be derived from the level crossing temperature. For instance, considering the denominator of $T_{\times_i}$ in eq.~\eqref{Tx_1+N} --- which assumes a light axion scenario by default --- we have the requirement $T_{\times_i} > 0 \Longrightarrow f_{A_i}<f_{a_0}, \, \forall i$. In other words, this is exactly the original requirement in eq.~\eqref{condition_fa_N+1} that the decay constants of all ALPs must uniformly be smaller than the decay constant of the QCD axion. Similarly, in the heavy axion scenario, we can obtain the corresponding reverse condition, $T_{\times_i} > 0 \Longrightarrow f_{A_i}>f_{a_0}, \, \forall i$.} 
In this context, smaller corresponds to the light QCD axion scenario, while larger corresponds to the heavy QCD axion scenario.
Any deviation from this condition, where some decay constants are smaller and others are larger, would inevitably compromise the maximal mixing. 
As described in section~\ref{sec_axion_mass_mixing_1+2_anisotropic}, the anisotropic $1+2$ mixing scenario can be transformed into two $1+1$ mixing scenarios, which respectively correspond to the light and heavy QCD axion scenarios.
Therefore, by adhering to these two stringent conditions, we can ensure the occurrence of maximal mixing among multiple axion mass eigenstates.
 
Furthermore, building upon our previous discussions, it is crucial to highlight a significant aspect regarding the transfer of axion energy density:
\begin{itemize}
{\it 
\item The energy transfer ultimately takes place solely between the two axions that possess the closest masses.
}
\end{itemize}
In the context of the $1+N$ mixing scenario, we observe the transfer processes: first, the transfer from the QCD axion $a_0$ to the ALP $A_1$; second, the transfer from $A_1$ to $a_0$, and subsequently from $a_0$ to the ALP $A_2$; third, the transfer from $A_2$ to $a_0$, and subsequently from $a_0$ to $A_3$; $\cdots$; and finally, the transfer from $A_N$ to the QCD axion.
This represents an extension of the $1+2$ mixing scenario presented in section~\ref{sec_axion_mass_mixing_1+2_isotropic} and differs from the $1+1$ mixing scenario discussed in section~\ref{sec_axion_mass_mixing_1+1}.
However, the conclusion drawn in section~\ref{sec_axion_mass_mixing_1+1} aligns with our current findings: axion energy transfer occurs exclusively between the two axions with the closest masses. 
 
\subsubsection{Heavy QCD axion scenario} 
 
The above discussions focus on the light QCD axion scenario.
In addition, if we consider the heavy QCD axion scenario where the decay constants of ALPs are all larger than that of the QCD axion,
\begin{eqnarray}
f_{A_i}>f_{a_0},\, \forall i\, ,
\end{eqnarray} 
then the matrix of domain wall numbers should be taken as 
\begin{eqnarray}
\mathfrak{n}_{(N+1)\times(N+1)}=
\left(
\begin{array}{cccccc}
1  & ~ 1 & ~ 1 & ~ 1 &~\cdots & ~ 1\\
0  & ~ 1 & ~ 0 & ~ 0 &~\cdots & ~ 0\\
0  & ~ 0 & ~ 1 & ~ 0 &~\cdots & ~ 0\\
0  & ~ 0 & ~ 0 & ~ 1 &~\cdots & ~ 0\\
\vdots  & ~\vdots & ~\vdots &~\vdots & ~\ddots & ~ \vdots\\
0  & ~ 0 & ~ 0 & ~ 0 & ~ \cdots & ~ 1
\end{array}
\right)\, ,
\end{eqnarray}
and the mass mixing matrix is given by
\begin{eqnarray}
\mathbf{M}^2=
\left(
\begin{array}{ccccc}
m_{a_0}^2 & ~ \dfrac{m_{a_0}^2 f_{a_0}}{f_{A_1}} & ~ \dfrac{m_{a_0}^2 f_{a_0}}{f_{A_2}} & ~\cdots &~ \dfrac{m_{a_0}^2 f_{a_0}}{f_{A_N}}\\
\dfrac{m_{a_0}^2 f_{a_0}}{f_{A_1}}   & ~ m_{A_1}^2+\dfrac{m_{a_0}^2 f_{a_0}^2}{f_{A_1}^2} & ~ \dfrac{m_{a_0}^2 f_{a_0}^2}{f_{A_1} f_{A_2}} & ~\cdots & ~\dfrac{m_{a_0}^2 f_{a_0}^2}{f_{A_1} f_{A_N}}\\
\dfrac{m_{a_0}^2 f_{a_0}}{f_{A_2}}   & ~ \dfrac{m_{a_0}^2 f_{a_0}^2}{f_{A_1} f_{A_2}} & ~ m_{A_2}^2+\dfrac{m_{a_0}^2 f_{a_0}^2}{f_{A_2}^2} & ~\cdots & ~\dfrac{m_{a_0}^2 f_{a_0}^2}{f_{A_2} f_{A_N}}\\
\vdots  & ~\vdots & ~\vdots  & ~\ddots & ~ \vdots\\
\dfrac{m_{a_0}^2 f_{a_0}}{f_{A_N}}   & ~ \dfrac{m_{a_0}^2 f_{a_0}^2}{f_{A_1} f_{A_N}} & ~ \dfrac{m_{a_0}^2 f_{a_0}^2}{f_{A_2} f_{A_N}} & ~ \cdots & ~ m_{A_N}^2+\dfrac{m_{a_0}^2 f_{a_0}^2}{f_{A_N}^2}
\end{array}
\right)\, ,
\end{eqnarray}
with the effective mixing matrices
\begin{eqnarray}
\mathbf{M}_i^2=
\left(\begin{array}{cc}
m_{a_0}^2  & ~ \dfrac{m_{a_0}^2 f_{a_0}}{f_{A_i}}\\
\dfrac{m_{a_0}^2 f_{a_0}}{f_{A_i}} & ~ m_{A_i}^2+\dfrac{m_{a_0}^2 f_{a_0}^2}{f_{A_i}^2}
\end{array}\right)\, .
\end{eqnarray}
Notice that the relation in eq.~\eqref{relation_ma_1+N} should also be satisfied here.
Further details regarding this heavy QCD axion scenario will not be discussed in this context.
Compared to the aforementioned discussions, the evolution of axions in this scenario is not expected to exhibit significant deviations.
Nevertheless, the modifications in axion energy density will be substantial and deserve attention.

\subsection{Quantitative analysis of maximal axion mixing}
\label{sec_quantitative_analysis}

In this subsection, we conduct a quantitative analysis of the maximal mixing.
Earlier, we defined maximal mixing as the scenario in which axions experience the largest degree of effective mixing.
However, this definition might be somewhat vague. 
Quantitatively, we can associate maximal mixing with the number of level crossings, denoted as $n_\times$.
Specifically, maximal mixing corresponds to the scenario with the highest number of level crossings.

For instance, in the $1+1$ mixing scenario presented in section~\ref{sec_axion_mass_mixing_1+1}, we have the number $n_\times=1$.
Since there is at most one level crossing, this scenario is inherently maximal.
In the $1+2$ mixing scenario, for the isotropic case in section~\ref{sec_axion_mass_mixing_1+2_isotropic}, we have $n_\times=2$, while for the anisotropic case in section~\ref{sec_axion_mass_mixing_1+2_anisotropic}, there are two situations with $n_\times=1$. 
Therefore, only the mixing in the isotropic case constitutes maximal mixing.
Finally, in the $1+N$ mixing scenario discussed in section~\ref{sec_axion_mass_mixing_1+N}, we make an {\it a priori} assumption that the decay constants of all ALPs are either smaller or larger than that of the QCD axion, and we have $n_\times=N$. 
Hence, the $1+N$ mixing scenario depicted represents maximal mixing.
Notice that this does not mean that all $1+N$ mixing scenarios exhibit maximal mixing.

We can therefore conclude that for the mixing of the QCD axion and $N$ ALPs, if the number of single level crossings is exactly $N$ ($n_\times=N$), the system exhibits maximal axion mixing.

Subsequently, we demonstrate how a given model may or may not exhibit maximal mixing.
Since there is currently no highly specific string axiverse model for the $1+N$ mixing scenario, we focus solely on the relationship between the axion decay constants and conduct a general analysis.

For simplicity, we consider the case involving the mixing of the QCD axion and ten ALPs. 
By default, we assume that the masses of the ALPs are all smaller than the zero-temperature mass of the QCD axion. 
The ratio of the decay constant of the ALP to that of the QCD axion is denoted as $\eta_i$. 
For instance, we consider a case where $\log_{10}\left(\eta_i\right)$ exhibits a random distribution within the range of $\left(-3,\, 3\right)$,
\begin{eqnarray}
\{2.46,-1.28,0.26,-0.37,-0.31,-2.68,0.23,-2.33,0.52,-0.25\}\, .
\end{eqnarray}
Under these circumstances, maximal mixing cannot occur. 
Instead, the system splits into two independent mixing scenarios: $1+6$ mixing and $1+4$ mixing with $n_\times=6$ and $n_\times=4$, respectively.
The matrices of domain wall numbers are given by
\begin{eqnarray}
\mathfrak{n}_{7\times7}=
\left(
\begin{array}{ccccccc}
1  & ~ 0 & ~ 0 & ~ 0 &~0 & ~ 0& ~ 0\\
1  & ~ 1 & ~ 0 & ~ 0 &~0 & ~ 0& ~ 0\\
1  & ~ 0 & ~ 1 & ~ 0 &~0 & ~ 0& ~ 0\\
1  & ~ 0 & ~ 0 & ~ 1 &~0 & ~ 0& ~ 0\\
1  & ~ 0 & ~ 0 & ~ 0 &~1 & ~ 0& ~ 0\\
1  & ~ 0 & ~ 0 & ~ 0 &~0 & ~ 1& ~ 0\\
1  & ~ 0 & ~ 0 & ~ 0 &~0 & ~ 0& ~ 1
\end{array}
\right)\, ,\quad 
\mathfrak{n}_{5\times5}=
\left(
\begin{array}{ccccc}
1  & ~ 1 & ~ 1 & ~ 1 & ~ 1\\
0  & ~ 1 & ~ 0 & ~ 0 & ~ 0\\
0  & ~ 0 & ~ 1 & ~ 0 & ~ 0\\
0  & ~ 0 & ~ 0 & ~ 1 & ~ 0\\
0  & ~ 0 & ~ 0 & ~ 0 &  ~ 1\\
\end{array}
\right)\, ,
\end{eqnarray}
corresponding to the light and heavy QCD axion scenarios, respectively.
Furthermore, consider another example where $\log_{10}\left(\eta_i\right)$ are taken as
\begin{eqnarray}
\{2.1,0.45,1.85,-0.93,2.01,1.12,2.86,-2.18,0.33,1.25\}\, .
\end{eqnarray} 
In this case, maximal mixing cannot occur either. 
We have $1+2$ mixing and $1+8$ mixing, corresponding to $n_\times=2$ and $n_\times=8$, respectively.
Then the matrices of domain wall numbers are given by
\begin{eqnarray}
\mathfrak{n}_{3\times3}=
\left(
\begin{array}{ccc}
1  & ~ 0 & ~ 0\\
1  & ~ 1 & ~ 0\\
1  & ~ 0 & ~ 1
\end{array}
\right)\, ,\quad 
\mathfrak{n}_{9\times9}=
\left(
\begin{array}{ccccccccc}
1  & ~ 1 & ~ 1 & ~ 1 &~1 & ~ 1& ~ 1& ~ 1& ~ 1\\
0  & ~ 1 & ~ 0 & ~ 0 &~0 & ~ 0& ~ 0& ~ 0& ~ 0\\
0  & ~ 0 & ~ 1 & ~ 0 &~0 & ~ 0& ~ 0& ~ 0& ~ 0\\
0  & ~ 0 & ~ 0 & ~ 1 &~0 & ~ 0& ~ 0& ~ 0& ~ 0\\
0  & ~ 0 & ~ 0 & ~ 0 &~1 & ~ 0& ~ 0& ~ 0& ~ 0\\
0  & ~ 0 & ~ 0 & ~ 0 &~0 & ~ 1& ~ 0& ~ 0& ~ 0\\
0  & ~ 0 & ~ 0 & ~ 0 &~0 & ~ 0& ~ 1& ~ 0& ~ 0\\
0  & ~ 0 & ~ 0 & ~ 0 &~0 & ~ 0& ~ 0& ~ 1& ~ 0\\
0  & ~ 0 & ~ 0 & ~ 0 &~0 & ~ 0& ~ 0& ~ 0& ~ 1
\end{array}
\right)
\, .
\end{eqnarray}
Therefore, maximal mixing requires that the decay constants of all ALPs be uniformly either smaller or larger than that of the QCD axion.

\section{Cosmological implications}%%%%%%%%%%%%%%%%%%%%%%%%%%
\label{sec_cosmological_implications}

In this section, we briefly discuss the potential cosmological implications of axion mass mixing in the string axiverse, including axion relic density, isocurvature fluctuations, dark energy, domain walls, gravitational waves, and primordial black holes. 
Notice that this is only a brief overview of the aforementioned cosmological effects, intended to provide readers with a general understanding.
More detailed discussions would require analysis based on specific models and are thus not further elaborated upon.

\subsection{Dark matter}

The most immediate effect of axion mass mixing is on the abundance of axion dark matter, a topic that has been extensively discussed in previous literature concerning two-axion mixing \cite{Ho:2018qur, Cyncynates:2021xzw, Murai:2023xjn, Li:2023xkn, Murai:2024nsp}.
As we mentioned earlier, based on their impact on the relic density of the QCD axion, different types of axion mass mixing can be categorized into the light QCD axion scenario \cite{Daido:2015cba, Li:2023uvt} and the heavy QCD axion scenario \cite{Cyncynates:2023esj}.
Notice that this classification also applies to ALP dark matter as well as to mixing scenarios involving more than two axions.

Specifically, in the string axiverse, the QCD axion with a higher-scale decay constant can be diluted by the late out-of-equilibrium decay of some heavy moduli \cite{Acharya:2010zx}.
Within this framework, the overproduction and underproduction of dark matter abundance of the QCD axion (or ALPs) can be naturally accounted for through their mass mixing.
As an example, let us discuss the abundance of QCD axion dark matter in the $1+2$ mixing scenario where eq.~\eqref{dw_matrix_1+2_light} is adopted.\footnote{For an example concerning the axion dark matter abundance in a $1+1$ mixing scenario, please refer to recent refs.~\cite{Li:2023xkn, Murai:2024nsp}.}
For simplicity, we consider the pre-inflationary scenario where the PQ symmetry is spontaneously broken during inflation, and the axion energy density transfers at the level crossing temperatures $T_{\times_i}$ are adiabatic.\footnote{Notice that the adiabatic condition is assumed throughout this work, which is a reasonable consideration given the large hierarchy of the axion decay constants. Nevertheless, non-adiabatic effects, such as those described by the generalized Landau-Zener mechanism, may become relevant during multiple level crossings.}
As shown in figure~\ref{fig_me_1+2_isotropic}, we observe three axion energy density transfer processes: first, the transfer from the QCD axion $a_0$ to the ALP $A_1$; second, the transfer from $A_1$ to $a_0$, and subsequently from $a_0$ to the ALP $A_2$; and third, the transfer from $A_2$ to the QCD axion.
Therefore, to determine the relic density of the QCD axion, we need to start with the ALP $A_2$, whose initial energy density at the oscillation temperature $T_{{\rm osc}, {A_2}}$ is given by
\begin{eqnarray}
\rho_{{A_2},{\rm osc}}=\frac{1}{2}m_{A_2}^2 f_{A_2}^2 \Theta_{{\rm i},A_2}^2\, ,
\end{eqnarray}
where $\Theta_{{\rm i},A_2}$ is the initial misalignment angle of $A_2$.
Then at $T_{\times_2}<T<T_{{\rm osc}, {A_2}}$, the energy density of $A_2$ is adiabatically invariant.
At the level crossing temperature $T_{\times_2}$, this energy density can be described by
\begin{eqnarray}
\rho_{A_2,\times_2}=\frac{1}{2}m_{A_2}^2 f_{A_2}^2 \Theta_{{\rm i},A_2}^2 \left(\frac{\mathfrak{a}_{{\rm osc},A_2}}{\mathfrak{a}_{\times_2}}\right)^3 \, ,
\end{eqnarray}
where $\mathfrak{a}_{{\rm osc},A_2}$ and $\mathfrak{a}_{\times_2}$ are the scale factors at $T_{{\rm osc},A_2}$ and $T_{\times_2}$, respectively.
Subsequently, $\rho_{A_2,\times_2}$ is transferred to the QCD axion, and the axion energy density is adiabatically invariant until the current cosmic microwave background (CMB) temperature $T_0$.
Consequently, we obtain the present QCD axion energy density
\begin{eqnarray}
\rho_{a_0,0}=\frac{1}{2}m_{a_0,0}m_{A_2} f_{A_2}^2 \Theta_{{\rm i},A_2}^2 \left(\frac{\mathfrak{a}_{{\rm osc},A_2}}{\mathfrak{a}_0}\right)^3 \, ,
\end{eqnarray}
where $\mathfrak{a}_0$ is the scale factor today.
Compared to the case without mass mixing (see also appendix~\ref{app_axion_relic_density_without_mixing}), the QCD axion relic density can be eﬀectively suppressed as
\begin{eqnarray}
\sim\left(\dfrac{m_{a_0,{\rm osc}}}{m_{A_2}}\right)^{1/2} \left(\dfrac{f_{A_2}}{f_{a_0}}\right)^2 \ll 1\, ,
\end{eqnarray}
where $m_{a_0,{\rm osc}}$ corresponds to the QCD axion mass at the oscillation temperature $T_{{\rm osc}, {a_0}}$.
Although this formula appears similar to eq.~(3.26) in a $1+1$ mixing scenario presented in ref.~\cite{Li:2023xkn}, it fundamentally differs from the mutual transfer between the QCD axion and ALP in that two-axion framework.
Notice that this relates to the light QCD axion scenario, which is valid when the QCD axion has a higher-scale decay constant. 

If the matrix of domain wall numbers outlined in eq.~\eqref{dw_matrix_1+2_heavy} is adopted, corresponding to the heavy QCD axion scenario, the axion abundance can be enhanced as
\begin{eqnarray}
\sim\left(\dfrac{m_{a_0,{\rm osc}}}{m_{A_2}}\right)^{1/2} \left(\dfrac{f_{A_2}}{f_{a_0}}\right)^2 \gtrsim 1\, ,
\end{eqnarray}
where $f_{A_2} > f_{a_0}$.
In addition, when extending the $1+2$ mixing scenario to the more general $1+N$ scenario, we find that the abundance of the QCD axion depends only on the ALP with the largest mass in the maximal mixing as
\begin{eqnarray}
\sim\left(\dfrac{m_{a_0,{\rm osc}}}{m_{A_N}}\right)^{1/2} \left(\dfrac{f_{A_N}}{f_{a_0}}\right)^2 \, ,
\end{eqnarray}
where eq.~\eqref{m_A_i_1+N} is adopted in this scenario.
Lastly, the abundance of ALPs can be analyzed using a similar method and will not be elaborated further here.\footnote{Very recently, the multi-component axion dark matter model corresponding to the $1+N$ mixing scenario proposed in this work was discussed in detail in ref.~\cite{Li:2025uwq}. In this framework, the dark matter is composed of the QCD axion and many ultra-light ALPs, where different axions can dominate after mixing depending on the relationships among their decay constants.}

\subsection{Isocurvature fluctuations}

Subsequently, another notable effect of axion mass mixing is its impact on isocurvature fluctuations, with scenarios involving two-axion mixing having been discussed in previous literature \cite{Kitajima:2014xla, Daido:2015cba}.
If the PQ symmetry is broken during inflation, the QCD axion will acquire quantum fluctuations, which lead to scale-invariant cold dark matter isocurvature density perturbations imprinted on the CMB temperature anisotropy.
Similarly, ALPs can also generate isocurvature perturbations. 
The amount of isocurvature perturbations is stringently constrained by CMB observations \cite{Planck:2018vyg, Planck:2018jri}.

In a multi-axion system, the quantum fluctuations of the QCD axion and ALPs are given by
\begin{eqnarray}
\delta\theta_{\rm i}=\dfrac{H_I}{2\pi f_{a_0}}\, , \quad \delta\Theta_{{\rm i},A_i}=\dfrac{H_I}{2\pi f_{A_i}}\, ,
\end{eqnarray}
where $H_I$ is the Hubble parameter during inflation.
Consequently, the dark matter isocurvature perturbation can be described by
\begin{eqnarray}
\Delta_{\mathcal{S}, \rm DM}=\left(\dfrac{\Omega_{\rm tot}}{\Omega_{\rm DM}}\right)\Delta_{\mathcal{S}, \rm axion}\, ,
\end{eqnarray} 
with
\begin{eqnarray}
\Omega_{\rm tot}=\Omega_{a_0}+\sum_{i=1}^N \Omega_{A_i}\, ,
\end{eqnarray}
where $\Omega_{\rm tot}$ and $\Omega_{\rm DM}$ are the density parameters for the total axions and the observed dark matter, respectively.
The power spectrum $\Delta_{\mathcal{S}, \rm axion}^2$ is given by 
\begin{eqnarray}
\Delta_{\mathcal{S}, \rm axion}^2=\left(\dfrac{\partial \ln \Omega_{\rm tot}}{\partial \theta_{\rm i}}\right)^2 \left(\dfrac{H_I}{2\pi f_{a_0}}\right)^2+\sum_{i=1}^N \left(\dfrac{\partial \ln \Omega_{\rm tot}}{\partial \Theta_{{\rm i},A_i}}\right)^2 \left(\dfrac{H_I}{2\pi f_{A_i}}\right)^2\, ,
\end{eqnarray} 
which can be rewritten as
\begin{eqnarray}
\bigg|\dfrac{\Delta_{\mathcal{S}, \rm axion}}{\delta\theta_{\rm i}}\bigg|^2=\left(\dfrac{\partial \ln \Omega_{\rm tot}}{\partial \theta_{\rm i}}\right)^2+\sum_{i=1}^N \left(\dfrac{\partial \ln \Omega_{\rm tot}}{\partial \Theta_{{\rm i},A_i}}\right)^2 \left(\dfrac{f_{a_0}}{f_{A_i}}\right)^2\, .
\end{eqnarray}
Due to the presence of mass mixing, the abundance of axions depends not only on the initial misalignment angle but also on the mass ratios among these axions.
As a result, the isocurvature perturbations of axions can be effectively altered.
Furthermore, under certain specific conditions, as illustrated in a case of two-axion mixing in ref.~\cite{Kitajima:2014xla}, they can be effectively suppressed.

\subsection{Dark energy}

Furthermore, the effect of axion mass mixing also has a potential impact on the composition of dark energy.
Recently, a case of two-axion mixing was investigated in refs.~\cite{Muursepp:2024mbb, Muursepp:2024kcg}.
Considering that ALP dominates the cosmological energy density and acts as dark energy, achieving the observed dark energy density requires an enhancement in the ALP number density through a specific mechanism.
Assuming the QCD axion comprises dark matter, a non-adiabatic level crossing between the QCD axion and an ALP shortly before matter-dark energy equality may convert a small fraction of dark matter into dark energy.
At the level crossing, in order to reach the matter-dark energy equality at $z_{\rm DE}\sim 5$, the ratio of dark energy density to matter density is required to be approximately \cite{Muursepp:2024mbb}
\begin{eqnarray}
\dfrac{\rho_{\rm DE}}{\rho_m}\bigg|_{T=T_\times}=\left(\dfrac{1+z_{\rm DE}}{1+z_\times}\right)^3 \sim 1\%-2\%\, ,
\end{eqnarray}
where $z$ represents the redshift.
This process provides a mechanism for boosting the number density of ALP and thus addresses this issue.

We anticipate that this mechanism will also be applicable to the $1+2$ mixing scenario, and furthermore, to the $1+N$ mixing scenario discussed previously.
However, given that the number of ALPs will be more than one in these cases, the specific effects are still unclear and require further exploration.

\subsection{Domain walls}

The formation of domain walls due to two-axion mixing has been investigated in refs.~\cite{Daido:2015bva, Lee:2024toz, Li:2024psa}. 
In this scenario, axion oscillations begin shortly before the level crossing temperature. 
If the initial energy density is sufficient to overcome the potential barrier, axion dynamics exhibit chaotic behavior, potentially leading to the formation of domain walls. 
Since cosmic strings do not form in this context, these domain walls remain stable on cosmological scales.

Given that domain walls evolve more slowly with cosmic expansion compared to radiation or matter, they will ultimately come to dominate the Universe, contradicting the standard cosmology. 
Therefore, to prevent the domain wall problem, they must annihilate before achieving dominance. 
Typically, a slight energy difference between different vacua can be introduced to facilitate the annihilation of domain walls. 
However, in the mixing scenario, the domain walls will naturally become unstable and annihilate due to an inherent bias term. 
Specifically, the evolution of domain walls over time can be expressed as \cite{Press:1989yh, Hindmarsh:1996xv, Garagounis:2002kt, Oliveira:2004he}
\begin{eqnarray}
\rho_{\rm wall}(t) = \mathcal{A} \frac{\sigma_{\rm wall}}{t}\, ,
\end{eqnarray}
where $\mathcal{A}$ represents the scaling parameter and $\sigma_{\rm wall}$ represents the tension of domain walls.
The annihilation of domain walls becomes significant when the tension pressure $p_T\simeq\rho_{\rm wall}$ is comparable to the volume pressure $p_V\simeq \Lambda_b^4$, yielding
\begin{eqnarray}
\mathcal{A}\frac{\sigma_{\rm wall}}{t_{\rm ann}} \simeq \Lambda_b^4\, ,
\end{eqnarray}
where $\Lambda_b$ is a bias parameter and $t_{\rm ann}$ is the domain walls annihilation time.

It is important to note that when considering mixing in scenarios involving more than two axions, the previously mentioned discussions regarding the formation of domain walls may become relatively complicated.
Additionally, the annihilation of domain walls must take place prior to the Big Bang Nucleosynthesis (BBN) epoch to avoid stringent constraints.
The BBN epoch is a fundamental period in the early Universe during which the lightest atomic nuclei were formed. 
If domain walls were to persist beyond this epoch, they would impose stringent and potentially insurmountable constraints on the theoretical models. 

\subsection{Gravitational waves}

Consequently, it is natural to consider the gravitational waves emitted by the annihilation of the aforementioned domain walls, which have also been extensively discussed in the literature \cite{Li:2024psa, Cyncynates:2022wlq, Chen:2021wcf, Kitajima:2023cek, Lee:2024xjb}.

In this case, the gravitational waves can be characterized by their peak frequency $f_p$ and peak amplitude $\Omega_{\rm GW}(t)_p$. 
The peak frequency corresponds to the Hubble parameter at the time of annihilation and is redshifted due to the cosmic expansion. 
The present peak frequency depends on several parameters, including the domain walls tension and energy scales.
The peak amplitude of gravitational waves at the annihilation time is given by a formula involving the domain walls tension and the efficiency parameter. 
The present peak amplitude is then calculated, taking into account the change in the number of effective degrees of freedom. 
Finally, the present gravitational waves spectrum is described by a piecewise function, with different power-law dependencies on frequency for low and high frequencies \cite{Hiramatsu:2013qaa}
\begin{eqnarray}
\Omega_{\rm GW}h^2=
\begin{cases}
\Omega_{\rm GW}(t_0)_ph^2\left(\dfrac{f}{f_{p,0}}\right)^3\, , &f \le f_{p,0}\\
\Omega_{\rm GW}(t_0)_ph^2\left(\dfrac{f_{p,0}}{f}\right)\, , &f > f_{p,0}
\end{cases} 
\end{eqnarray} 
where $h$ represents the reduced Hubble parameter, $f_{p,0}$ and $\Omega_{\rm GW}(t_0)_p$ represent the present peak frequency and amplitude, respectively. 

In a two-axion mixing scenario \cite{Li:2024psa}, we predict the gravitational wave spectra with the peak frequency of approximately $0.2\, \rm nHz$ and the peak amplitude of around $5\times 10^{-9}$, which can be tested by future pulsar timing array (PTA) projects. 

\subsection{Primordial black holes} 

Similarly, primordial black holes can also form from the collapse of these domain walls, and even result in the formation of supermassive black holes \cite{Li:2024psa, Cyncynates:2022wlq, Chen:2021wcf}. 
Closed domain walls can shrink to the Schwarzschild radius during their annihilation and subsequently collapse into primordial black holes \cite{Ferrer:2018uiu, Ge:2023rrq}. 
The collapse of these domain walls is considered to occur in an approximately spherically symmetric manner \cite{Gelmini:2022nim, Gelmini:2023ngs, Gouttenoire:2023gbn}. 
Specifically, the fraction of primordial black holes can be described by\footnote{Notice that this formula, along with its corresponding parameters, applies only to two-axion mixing case; extending it to $1+2$ and even $1+N$ mixing scenarios requires further investigation.} \cite{Li:2024psa}
\begin{eqnarray}
f_{\rm PBH}\simeq 4\left(\dfrac{45 m_{\rm Pl}^6}{64 \pi^3}\right)^{(\alpha+1)/4} \dfrac{g_{*s}(T_0)}{g_*(T_0)}\dfrac{g_*(T_f)^{(3-\alpha)/4}}{g_{*s}(T_f)}\dfrac{M_{\rm PBH}^{-(\alpha+1)/2}}{T_{\rm ann}^\alpha T_0} \dfrac{\rho_R(T_0)}{\rho_{\rm DM}(T_0)}
 \, ,
\end{eqnarray} 
where $m_{\rm Pl}$ is the Planck mass, $\alpha$ is a positive parameter \cite{Kawasaki:2014sqa}, $T_{\rm ann}$ is the domain walls annihilation temperature, $T_f$ is the primordial black holes formation temperature, $M_{\rm PBH}$ is the primordial black holes mass, $\rho_R$ and $\rho_{\rm DM}$ are the radiation and dark matter energy densities, respectively. 

We also demonstrate that primordial black holes, within a mass range of $\mathcal{O}(10^5-10^8)\, M_\odot$, could potentially form in a two-axion mixing scenario, accounting for approximately $10^{-5}$ of the cold dark matter \cite{Li:2024psa}. 
Furthermore, two-axion mixing has an impact on another formation mechanism of primordial black holes \cite{Kitajima:2020kig, Li:2023det}, as well as on the formation of supermassive black holes at high redshifts \cite{Li:2023zyc}.
 
\section{Conclusion and outlook}%%%%%%%%%%%%%%%%%%%%%%%%%%%Conclusion 
\label{sec_Conclusion}

In summary, we have investigated axion mass mixing in the string axiverse and have adopted a bottom-up perspective to determine what conditions an axion model must fulfill in order to exhibit maximal mixing.
Our exploration began with an overview of axions in string theory, encompassing both model-independent and model-dependent axions. 
We then transitioned to a review of the type IIB string axiverse, emphasizing the significance of the LVS in realizing an axiverse potential.
Within this context, we have presented several specific LVS axiverse models, such as the one featuring one QCD axion plus one ALP, one QCD axion plus two ALPs, and even more complex scenarios involving one QCD axion plus multiple ALPs. 
These models provide a theoretical basis for our mixing scenarios of $1+1$, $1+2$, and $1+N$ ($N > 2$) configurations, where $1$ denotes the number of the QCD axion and $N$ signifies the number of ALPs, with each scenario featuring unique axion masses and decay constants. 

Our investigation of axion mass mixing revealed crucial insights. 
Through detailed analysis, we have derived the mass mixing matrices and their corresponding eigenvalues, which have enabled us to understand the complex dynamics of axion mass eigenstates in the mixing. 
Notably, we found that maximal mixing among multiple axion mass eigenstates occurs under specific conditions: the masses of all ALPs must be smaller than the mass of the QCD axion, with no two ALP masses being equal, and the decay constants of all ALPs must uniformly be either smaller or larger than the decay constant of the QCD axion. 
We also found that the transfer of axion energy density ultimately only occurs between the two axions with the closest masses.
These insights lay the groundwork for understanding axion dynamics within the string axiverse and are also applicable to generalized multi-axion mixing scenarios.

Moreover, we delved into the cosmological implications of axion mass mixing, including its potential roles in dark matter, isocurvature fluctuations, dark energy, domain walls, gravitational waves, and primordial black holes. 
The rich phenomenology, particularly within the context of the string axiverse, offers intriguing possibilities for addressing some of the most pressing questions in modern cosmology.

The string axiverse presents a fascinating and complex landscape for axion physics. 
Our study has provided a detailed examination of its various facets, from axion models and mass mixing to cosmological implications. 
As we continue to unravel the mysteries of the Universe, the axiverse remains a promising arena for theoretical and observational exploration. 
Future work in this area will deepen our understanding of axions and their potential roles within the broader context of early Universe cosmology.

\section*{Acknowledgments}%%%%%%%%%%%%%%%%Acknowledgments

We thank Michele Cicoli for useful discussions.
H.J.L. was supported in part by the Institute of Theoretical Physics, CAS, and the International Centre for Theoretical Physics Asia-Pacific.
Y.F.Z. was supported by the CAS Project for Young Scientists in Basic Research YSBR-006, the National Key R\&D Program of China (Grant No.~2017YFA0402204), and the National Natural Science Foundation of China (NSFC) (Grants No.~11821505, No.~11825506, and No.~12047503).

%%%%%%%%%%%%%%%%%%%%%%%%%%%%%
\appendix

\section{Divisor volumes}
\label{app_divisor_volumes}

Consider the orientifold projections $h_-^{1,1}=0\to h^{1,1}=h_+^{1,1}$, and defining $\{\hat{D}_i\}_{i=1}^{h^{1,1}}$ as a basis of $H^{1,1}(X, \mathbb{Z})$, the Kähler form can be described by
\begin{eqnarray}
J= t^i \hat{D}_i \, ,
\end{eqnarray}
where $t^i$ represent the real coordinates on the Kähler moduli space, and each two-form $\hat{D}_i$ is Poincar$\acute{\rm e}$ dual to the divisor $D_i$.
The volume of the internal manifold is given by
\begin{eqnarray}
\mathcal{V}=\dfrac{1}{3!}\int_X J\wedge J\wedge J=\dfrac{1}{3!}k_{ijk} t^i t^j t^k\, ,
\end{eqnarray}
where $k_{ijk}$ are the triple intersection numbers
\begin{eqnarray}
k_{ijk}=\int_X \hat{D}_i\wedge \hat{D}_j\wedge \hat{D}_k=D_i\cdot D_j\cdot D_k \, .
\end{eqnarray}
The natural holomorphic coordinates appearing in type IIB flux compactifications are the Einstein-frame volumes of the divisor
\begin{eqnarray}
\tau_i=\dfrac{1}{2}\int_X \hat{D}_i\wedge J\wedge J=\dfrac{\partial \mathcal{V}}{\partial t^i}=\dfrac{1}{2}k_{ijk}t^j t^k\, .
\end{eqnarray}
This transformation of coordinates amounts to a Legendre transform on $\mathcal{V}$.
The Kähler potential is determined as
\begin{eqnarray}
\mathcal{K}=\dfrac{K}{M_{\rm Pl}^2}=-2\ln \mathcal{V}\, , \quad \mathcal{K}_{ij}=\dfrac{\partial \mathcal{K}}{\partial \tau_i \partial \tau_j}\, , \quad \mathcal{K}^{ij}=\tau_i \tau_j -k_{ijk}t^k \mathcal{V}\, .
\end{eqnarray}

\section{Axion couplings to gauge fields}
\label{app_axion_couplings}

Axion couplings to gauge field strengths originates from the KK reduction of the Chern-Simons (CS) term in the D-brane action.
Considering the axion fields $c^a$ and $c_\alpha$, the effective Lagrangian is given by \cite{Cicoli:2012sz}
\begin{eqnarray}
\begin{aligned}
\mathcal{L}&\supset-\dfrac{e^{\Phi}}{4\mathcal{V}^2}\left(d c^a+\dfrac{M_{\rm Pl}}{2\pi} e^{-\Phi} \mathcal{A}_i r^{ia}\right)\mathcal{K}^{ab}\wedge \star \left(d c^b+\dfrac{M_{\rm Pl}}{2\pi} e^{-\Phi} \mathcal{A}_j r^{jb}\right)\\
&-\dfrac{1}{8}\left(d c_\alpha+\dfrac{M_{\rm Pl}}{\pi} \mathcal{A}_i q_{i\alpha}\right)\mathcal{K}^{\alpha\beta}\wedge \star \left(d c_\beta+\dfrac{M_{\rm Pl}}{\pi} \mathcal{A}_j q_{j\beta}\right)\\
&+\dfrac{M_{\rm Pl}^2}{4\pi^2 \mathcal{V}^2} e^{-\Phi} \mathcal{A}_i \mathcal{A}_j r^{ia} r^{jb} \mathcal{K}^{ab} +\dfrac{M_{\rm Pl}^2}{8\pi^2} \mathcal{A}_i \mathcal{A}_j q_{i\alpha} q_{j\beta} \mathcal{K}_{\alpha\beta}\\
&+\dfrac{1}{4\pi M_{\rm Pl}}\left(q_{ia}c^a+r^{i\alpha}c_\alpha\right){\rm Tr}(F\wedge F)- \dfrac{r^{i\alpha}\tau_\alpha}{4\pi M_{\rm Pl}}{\rm Tr}(F_i\wedge \star F_i)  \, ,
\end{aligned}
\end{eqnarray}
where $\Phi$ represents the axio-dilaton, $\mathcal{A}_i$ are the $\rm U(1)$ gauge bosons, $r^{ia}$ and $q_{i\alpha}$ are the couplings of the two-forms.
Notice that the orientifold-odd fields $c^a$ have decay constants that are equal to or larger than the string scale, rendering them negligible. 
Therefore, we focus on the case where $h_-^{1,1}=0$.

\section{More details about the type IIB string axiverse}
\label{app_type_IIB_string_axiverse}

The type IIB string axiverse exhibits two distinct regimes, depending on the VEV of the tree-level superpotential $W_0$.
In the ample divisor case where $W_0\ll 1$, the single non-perturbative effect is capable of generating a minimum for all the Kähler moduli, lifting only one axion while leaving $h^{1,1}-1$ axions massless at leading order \cite{Acharya:2010zx}.
The main challenges in realizing this scenario microscopically lie in finding a rigid ample divisor and avoiding chiral intersections via gauge flux selection.

On the other hand, in the case where $W_0$ is of order one, the LVS leads to an axiverse with one heavy axion and multiple light axions \cite{Balasubramanian:2005zx, Blumenhagen:2009gk, Cicoli:2008va}. 
The heavy axion arises from a del Pezzo divisor $T_{\rm dP}$ fixed by non-perturbative effects, while the light axions are stabilized perturbatively by the $\alpha'$ and $g_s$ effects, and acquire a potential from higher-order instanton effects.
Notice that some light axions could be eaten up by massive $\rm U(1)$ gauge bosons via the Stückelberg mechanism.
In this context, the scalar potential comprises various contributions that scale differently with the exponentially large total volume \cite{Cicoli:2012sz, Cicoli:2011qg}
\begin{eqnarray}
V=V_D+V_F^{\rm tree}+V_F^{\rm np}+V_F^{\rm p}\, ,
\end{eqnarray}
where $V_D$ represents the D-term potential that scales as $V_D\sim\mathcal{O}(\mathcal{V}^{-2})$, $V_F^{\rm tree}$ represents the tree-level F-term scalar potential that scales as $V_F^{\rm tree}\sim\mathcal{O}(\mathcal{V}^{-2})$, $V_F^{\rm np}$ represents the non-perturbative scalar potential that scales as $V_F^{\rm np}\sim\mathcal{O}(\mathcal{V}^{-3})$, and $V_F^{\rm p}$ represents the perturbative potential from the $\alpha'$ and $g_s$ corrections to the Kähler potential.

The D-term conditions stabilize $d$ combinations of Kähler moduli, resulting in $n_{\rm ax}=h^{1,1}-d-1$ flat directions. 
Consequently, the $d$ combinations of axions are eaten up, leaving only $n_{\rm ax}$ light axions. 
Furthermore, this requires that $n_{\rm ax}\ge 2$ to ensure the existence of a QCD axion candidate along with at least one additional ALP.
The non-perturbative scalar potential $V_F^{\rm np}$ arises from corrections to the superpotential $W$, which stabilize $T_{\rm dP}$ at a small size and render the corresponding axions massive, with $m_{a_{\rm dP}}\sim m_{\tau_{\rm dP}}$ \cite{Conlon:2005ki}.
The perturbative potential $V_F^{\rm p}$ lifts the remaining flat directions from the D-terms and non-perturbative effects, leaving all corresponding axions massless. 
The $\alpha'$ correction leads to a scalar potential that scales as $V_F^{\alpha'}\sim\mathcal{O}(\mathcal{V}^{-3})$ \cite{Becker:2002nn}, which yields an exponentially large internal volume \cite{Balasubramanian:2005zx}
\begin{eqnarray}
\mathcal{V}\sim W_0 e^{a\tau_{\rm dP}}\, ,
\end{eqnarray}
with a light volume mode mass $m_\mathcal{V}\sim W_0 M_{\rm Pl}/\mathcal{V}^{3/2}\sim \sqrt{m_{3/2}^3/M_{\rm Pl}}$, where $a=2\pi/n$, and $m_{3/2}$ is the gravitino mass.
The $g_s$ correction is sub-leading and leads to a scalar potential that scales as $V_F^{g_s}\sim\mathcal{O}(\mathcal{V}^{-(3+p)})$ with $p>0$ \cite{Cicoli:2007xp}.
Then the sub-leading higher-order non-perturbative effects slightly lift the $n_{\rm ax}$ axion directions, resulting in the emergence of the LVS axiverse \cite{Cicoli:2012sz}
\begin{eqnarray}
V_{W_{\rm np}}(c_i)\simeq -\sum_{i=1}^{n_{\rm ax}}\dfrac{n_i a_i \tau_i W_0}{\mathcal{V}^2}e^{-n_i a_i \tau_i}\cos\left(n_i a_i c_i\right)\, .
\end{eqnarray}
In addition, axions can acquire mass through non-perturbative corrections to the Kähler potential. 
These corrections can only dominate the QCD axion potential if there is a single instanton superpotential contribution \cite{Conlon:2006tq}.
In this case, the ALP masses are expected to be smaller than that of the QCD axion.

\section{Derivation of the level crossing temperature}
\label{app_level_crossing_temperature}

Here we present the derivation of the level crossing temperature in eq.~\eqref{Tx_1+1} as an example.
The temperature-dependent QCD axion mass can be described by
\begin{eqnarray}
m_{a_0}(T)=
\begin{cases}
m_{a_0,0}\, , & T\leq T_{\rm QCD}\\ 
m_{a_0,0}\left(\dfrac{T}{T_{\rm QCD}}\right)^{-b}\, , & T> T_{\rm QCD} 
\end{cases} 
\label{mQCDT}
\end{eqnarray} 
where $m_{a_0,0}$ is the zero-temperature mass of the QCD axion \cite{GrillidiCortona:2015jxo}, $T_{\rm QCD}\simeq150\, \rm MeV$ is the QCD phase transition critical temperature, and $b\simeq4.08$ is an index taken from the dilute instanton gas approximation \cite{Borsanyi:2016ksw}.
Substituting eq.~\eqref{mass_eigenvalues} into eq.~\eqref{differential_equation}, we obtain
\begin{eqnarray}
\dfrac{d\left(m_h^2(T)-m_l^2(T)\right)}{dT}\bigg|_{T>T_{\rm QCD}}&=&0\, , \\
-\dfrac{2b m_{a_0,0}^2(\frac{T}{T_{\rm QCD}})^{-4b}\left(m_{A_1}^2 f_{A_1}^2(\frac{T}{T_{\rm QCD}})^{2b}+f_{a_0}^2\left(m_{a_0,0}^2-m_{A_1}^2(\frac{T}{T_{\rm QCD}})^{2b}\right)\right)}{T\sqrt{-4m_{a_0,0}^2(\frac{T}{T_{\rm QCD}})^{-2b}m_{A_1}^2 f_{a_0}^4+\left(m_{A_1}^2 f_{A_1}^2+f_{a_0}^2\left(m_{A_1}^2+m_{a_0,0}^2(\frac{T}{T_{\rm QCD}})^{-2b}\right)\right)^2}}&=&0\, ,\quad\quad~\\
&\Downarrow& \nonumber\\
-2b m_{a_0,0}^2\left(-m_{A_1}^2 f_{A_1}^2+f_{a_0}^2\left(-m_{a_0,0}^2\left(\frac{T}{T_{\rm QCD}}\right)^{-2b}+m_{A_1}^2\right)\right)&=&0\, ,\\
-m_{A_1}^2 f_{A_1}^2+f_{a_0}^2\left(-m_{a_0,0}^2\left(\frac{T}{T_{\rm QCD}}\right)^{-2b}+m_{A_1}^2\right)&=&0\, .
\end{eqnarray}
Subsequently, it is straightforward to obtain
\begin{eqnarray}
\left(\frac{T}{T_{\rm QCD}}\right)^{-2b}=\dfrac{m_{A_1}^2-\dfrac{m_{A_1}^2 f_{A_1}^2}{f_{a_0}^2}}{m_{a_0,0}^2}\, ,
\end{eqnarray}
which further yields the level crossing temperature
\begin{eqnarray}
T\equiv T_\times= T_{\rm QCD} \left(\dfrac{m_{a_0,0}^2 f_{a_0}^2}{m_{A_1}^2 \left(f_{a_0}^2-f_{A_1}^2\right)}\right)^{\frac{1}{2b}} \, . 
\end{eqnarray}
The level crossing temperatures for other cases can be derived similarly.

\section{Axion relic density without mass mixing}
\label{app_axion_relic_density_without_mixing}

Here we show the axion relic density without mass mixing via the misalignment mechanism.
Notice that we just consider the pre-inflationary scenario where the PQ symmetry is spontaneously broken during inflation.

At high cosmic temperatures $T\gg T_{\rm QCD}$, the QCD axion field remains frozen at its initial misalignment angle $\theta_{\rm i}$.
As the temperature decreases, it begins to oscillate when its mass becomes comparable to the Hubble parameter.
The oscillation temperature $T_{{\rm osc}, {a_0}}$ is given by
\begin{eqnarray}
m_{a_0}(T_{{\rm osc}, {a_0}})=3H(T_{{\rm osc}, {a_0}})\, ,
\end{eqnarray}
with the Hubble parameter
\begin{eqnarray}
H(T)=\sqrt{\dfrac{\pi^2 g_*(T)}{90}}\dfrac{T^2}{M_{\rm Pl}}\, .
\end{eqnarray} 
The axion initial energy density at $T_{{\rm osc}, {a_0}}$ is
\begin{eqnarray}
\rho_{a_0,{\rm osc}}=\dfrac{1}{2}m_{a_0,{\rm osc}}^2 f_{a_0}^2 \theta_{\rm i}^2\, ,
\end{eqnarray}
where $m_{a_0,{\rm osc}}$ is the axion mass at $T_{{\rm osc}, {a_0}}$.
For temperatures $T_0<T<T_{1,a}$, the QCD axion energy density is adiabatically invariant with the comoving number 
\begin{eqnarray}
N_{a_0} \equiv \dfrac{\rho_{a_0}}{m_{a_0}} \mathfrak{a}^3\, ,
\end{eqnarray}
where $\mathfrak{a}$ is the scale factor.
Using 
\begin{eqnarray}
\dfrac{\rho_{a_0,{\rm osc}} \mathfrak{a}_{{\rm osc},a_0}^3}{m_{a_0,{\rm osc}}}=\dfrac{\rho_{a_0,0} \mathfrak{a}_0^3}{m_{a_0,0}}\, ,
\end{eqnarray}
we can derive the QCD axion energy density at present
\begin{eqnarray}
\rho_{a_0,0} =\dfrac{1}{2} m_{a_0,0} m_{a_0,{\rm osc}} f_{a_0}^2 \theta_{\rm i}^2 \left(\dfrac{\mathfrak{a}_{{\rm osc},a_0}}{\mathfrak{a}_0}\right)^3\, ,
\end{eqnarray}
where $\mathfrak{a}_{{\rm osc},a_0}$ and $\mathfrak{a}_0$ are the scale factors at $T_{{\rm osc}, {a_0}}$ and $T_0$, respectively.
More precisely, the current QCD axion dark matter abundance is given by \cite{Li:2023xkn}
\begin{eqnarray}
\Omega_{a_0}h^2\simeq0.14 \left(\dfrac{g_{*s}(T_0)}{3.94}\right)\left(\dfrac{g_*(T_{{\rm osc}, {a_0}})}{61.75}\right)^{-0.42}\left(\dfrac{f_{a_0}}{10^{12}\, \rm GeV}\right)^{1.16}\left\langle\theta_{\rm i}^2f(\theta_{\rm i})\right\rangle\, ,
\end{eqnarray}
where $f(\theta_{\rm i})$ is the anharmonicity factor \cite{Lyth:1991ub, Visinelli:2009zm}.
In order to explain the observed dark matter abundance, we require an $\sim\mathcal{O}(1)$ initial misalignment angle
\begin{eqnarray}
\theta_{\rm i}\simeq0.87\left(\dfrac{g_{*s}(T_0)}{3.94}\right)^{-1/2}\left(\dfrac{g_*(T_{{\rm osc}, {a_0}})}{61.75}\right)^{0.21}\left(\dfrac{f_{a_0}}{10^{12}\, \rm GeV}\right)^{-0.58}\, .
\end{eqnarray}
Similarly, for the single-field ALP dark matter with a constant mass $m_{A_i}$, its abundance can be expressed as \cite{Li:2023xkn}
\begin{eqnarray}
\begin{aligned}
\Omega_{A_i} h^2&\simeq0.95 \left(\dfrac{g_{*s}(T_0)}{3.94}\right)\left(\dfrac{g_*(T_{{\rm osc},A_i})}{61.75}\right)^{-1}\\
&\times \left(\dfrac{m_{A_i}}{1\, \rm eV}\right)^{1/2}\left(\dfrac{f_{A_i}}{10^{12}\, \rm GeV}\right)^2\left\langle\Theta_{{\rm i},A_i}^2f(\Theta_{{\rm i},A_i})\right\rangle\, .
\end{aligned}
\end{eqnarray}

\bibliographystyle{JHEP}
\bibliography{references}

\end{document}